**Transcranial Focused Ultrasound for Identifying the Neural Substrate of Conscious Perception**

Daniel K. Freeman[1], Brian Odegaard[2], Seung-Schik Yoo[3], Matthias Michel[4]

[1] Massachusetts Institute of Technology, Lincoln Laboratory

[2] Department of Psychology, University of Florida

[3] Department of Radiology, Brigham and Women's Hospital, Harvard Medical School

[4] Department of Linguistics and Philosophy, Massachusetts Institute of Technology

**Abstract**

Identifying what aspects of brain activity are responsible for conscious perception remains one of the most challenging problems in science. While progress has been made through psychophysical studies employing EEG and fMRI, research would greatly benefit from improved methods for stimulating the brain in healthy human subjects. Traditional techniques for neural stimulation through the skull, including electrical or magnetic stimulation, suffer from coarse spatial resolution and have limited ability to target deep brain structures with high spatial selectivity. Over the past decade, a new tool has emerged known as transcranial focused ultrasound (tFUS), which enables the human brain to be stimulated safely and non-invasively through the skull with millimeter-scale spatial resolution, including cortical as well as deep brain structures. This tool offers an exciting opportunity for breakthroughs in consciousness research. Given the extensive preparation and regulatory approvals associated with tFUS testing, careful experimental planning is essential. Therefore, our goal here is to provide a roadmap for using tFUS in humans for exploring the neural substrate of conscious perception.

## Introduction

Understanding the relationship between brain activity and conscious experience remains one of the most challenging research problems in modern science. Despite the seemingly intractable nature of the problem, the field has grown considerably in recent years (Michel et al. 2019). In particular, there has been success in focusing on conscious perception of sensory stimuli as a stepping stone towards understanding consciousness more broadly (Mehta and Mashour 2013). This focus on perception offers an experimentally tractable strategy in which the stimulus can be controlled, brain activity can be monitored, and human subjects can report their sensations. These types of studies have continued to improve our understanding of the brain structures and neural circuits involved in conscious perception (Kronemer et al. 2022; Dykstra, Cariani, and Gutschalk 2017).

However, one of the challenges of these studies is that a given sensory input, such as a visual or tactile stimulus, can elicit activity in many different brain areas, making it difficult to establish which areas contribute directly to conscious perception, as opposed to areas playing a subconscious role in sensory processing. For example, a visual stimulus may evoke neural activity that correlates with conscious perception, but this does not imply that this neural activity constitutes the resulting conscious percept. Teasing out which neural structures can directly elicit conscious perception versus those that do not is still a central challenge of consciousness research.

To go beyond correlational findings and provide causal evidence, one promising approach is to directly stimulate specific brain regions through the skull, which could allow us to evaluate the role of specific brain regions. For example, transcranial magnetic stimulation (TMS) and transcranial alternating current stimulation (tACS) have been used in consciousness research (Bachmann 2018; Bareham et al. 2021; Cabral-Calderin and Wilke 2020; Raccah, Block, and Fox 2021). However, the spatial resolution of these stimulation techniques is relatively coarse, and they also have limited use in targeting deep brain structures. This puts limitations on the ability to selectively probe small targets, such as a region of visual cortex corresponding to a small region of the visual field, or millimeter-scale nuclei in the basal forebrain.

More recently, transcranial focused ultrasound (tFUS) has emerged as a powerful new tool for non-invasive brain stimulation (Darmani et al. 2022). tFUS has been shown to be safe (Lee et al. 2021) and is actively being tested for a number of clinical indications (Matt et al. 2024). Furthermore, tFUS has been used successfully in humans to target sensory cortex (Lee et al. 2015, 2016) as well as deep brain structures, including the nucleus accumbens (Mahoney et al. 2023), ventral posterolateral thalamus (Kim et al. 2023), and anterior nucleus of the thalamus (Fan et al. 2024).

Given the advantages of tFUS over other stimulation techniques, we believe it has great potential for use in consciousness research. However, there are significant barriers to entry in performing in-human tFUS testing for a number of reasons: (1) the high cost of commercially available systems, (2) the use of MRI and CT scans for image-guidance and in-situ acoustic power estimation, (3) knowledge of appropriate stimulation parameters, and (4) the needs for regulatory and safety approvals, for example, by Institutional Review Boards (IRB). With this review, we aim to lower the barrier to entry so that researchers in the field of consciousness can participate in tFUS testing, either independently or in collaboration with tFUS laboratories.

Furthermore, to help guide this research, we will describe potential tFUS experiments that can be performed to address open questions, such as: what is the role of prefrontal cortex in conscious perception (Michel 2022; Odegaard, Knight, and Lau 2017; Boly et al. 2017; Kozuch 2024)? Does conscious perception arise from global processing across many brain areas (Mashour et al. 2020) or is it generated within some local brain region (Malach 2021)? If it is generated globally, then how is activity in distant brain regions bound together to produce a unified percept (Stanislas Dehaene and Changeux 2011)? Can tFUS be used to test various theories of consciousness, such as Global Workspace or Integrated Information Theory (Storm et al. 2024; Seth and Bayne 2022)? What is the role of subcortical structures in generating conscious perception (Nieder 2021; Zheng 2024; Merker 2007)?

This review consists of three sections. First, we discuss the principles of transcranial ultrasound, highlighting results from human studies on sensory perception. Second, we will provide a brief overview of existing tools for studying conscious perception, including psychophysical studies with brain monitoring and direct stimulation of the brain, either non-invasively or during surgery. Third, we present a roadmap for potential tFUS experiments to address gaps not easily filled with existing tools (e.g., TMS). Our aim is to accelerate the adoption of tFUS in consciousness research to help identify the brain structures and neural circuits that are required for conscious perception.

**Transcranial Ultrasound for Neuromodulation**

Throughout the 20th century, developments in ultrasound technology resulted in commercial applications involving diagnostic imaging, as well as high-intensity ultrasound to thermally ablate tissue and break kidney stones (Bachu et al. 2021). However, the use of low-intensity ultrasound for neuromodulation has only recently been the focus of intense study, despite the observation as far back as the 1950's that neuronal firing can be reversibly suppressed in animals with ultrasound (Fry, Ades, and Fry 1958). Focused ultrasound is also being used for opening the blood-brain barrier for drug delivery (Cammalleri et al. 2020), but this technique will not be the focus of this review.

The resurgence in interest in low-intensity ultrasound for neuromodulation was the result of pioneering work from William Tyler, first in mouse hippocampal slices and ex-vivo whole brains (Tyler et al. 2008), and then transcranially in anesthetized mice (Tufail et al. 2010), followed by the first in-vivo demonstration of bimodal modulatory potentials (i.e., excitation and suppression) by Yoo and colleagues (Yoo et al., 2011). These animal studies enabled the first human tests with transcranial ultrasound by several groups starting in 2013 (Hameroff et al. 2013; Legon et al. 2014; Lee et al. 2015). For a comprehensive description of testing to date, including animal testing and ongoing clinical trials, please see (Pellow, Pichardo, and Pike 2024; Matt et al. 2024).

For researchers interested in applying tFUS to consciousness research, there are several relevant considerations, including the nature of the neural response (e.g., spatial and temporal resolution), a description of tFUS systems that are commonly used, and finally, a description of various methods for monitoring brain activity during tFUS testing. We will discuss these topics below.

*The Neural Response to Ultrasonic Stimulation*

Non-invasive ultrasonic stimulation of the brain involves a pressure wave being delivered through the skull. These time-varying pressure waves cause vibrations in neural tissue, and this can result in modulations of electrical activity. The ultrasonic pressure wave is usually focused such that the pressure will exceed some threshold for neural activity only within a predefined region of interest; this technique is referred to as transcranial focused ultrasound (tFUS). Importantly, this focal region can reach deep in the brain, allowing subcortical structures to be targeted. The spatial resolution that can be achieved is fundamentally limited by the frequency of the ultrasonic wave, with higher frequencies (shorter wavelengths) resulting in smaller focal regions. However, since higher frequencies result in more attenuation by the skull (~ 20 dB/cm/MHz), frequencies below 1 MHz are generally used. To a first-order, the smallest achievable focal spot size is approximately equal to the wavelength. For example, assuming the speed of sound in biological tissue is 1,500 m/s, then a frequency of 500 kHz will have a wavelength of 3 mm.

In practice, the spatial resolution for neuromodulation tends to be larger than the wavelength for a number of reasons. One major contributor to the spatial resolution is the transducer design; in particular, the relative size of the focal length and aperture size, referred to as the f-number. As with any focusing system, including optical or radio waves, if the focal length is small relative to the aperture, then the axial direction of the focal area tends to be elongated. Simulations and benchtop testing with skulls show elongation distances that exceed the wavelength considerably (e.g., tens of millimeters in length) (Yoon et al. 2018). This is particularly relevant for cortical targets, which are shallow relative to the scalp. Another factor that degrades spatial resolution is due to aberrations in the skull that can cause the 3D spatial pattern of pressure to become distorted, which can impact both the shape of the pressure field and location of the peak pressure. For this reason, a CT is often taken before tFUS testing in order to map variations in thickness across the skull. The CT data can be input to numerical simulations of the pressure field in the brain, thereby allowing the transducer to be adjusted to correct for the effects of the skull (e.g., by electronic steering of the beam with a phased-array). Another artifact associated with imperfect focusing is the presence of side lobes, which are periodic peaks in pressure located laterally relative to the peak. The amplitude of these side lobes depends strongly on the transducer design. This effect can be mitigated substantially with phased-arrays containing hundreds of ultrasonic transducer elements (Chaplin, Phipps, and Caskey 2018). Taken together, these effects that degrade spatial resolution may explain why sonication of the primary visual cortex in humans with a spherical transducer produces a diffuse perception of light rather than a localized phosphene (Lee et al. 2016). As discussed below, improvements in hardware, such as 100+ element phased-arrays and acoustic lenses, can address many of these issues, potentially bringing spatial resolution closer to the diffraction limit (e.g., ~3 mm spot size for a frequency of 500 kHz).

One of the primary challenges to optimizing the sonication parameters in tFUS is a lack of understanding of the mechanisms underlying the neuronal response. Thermal effects are likely not the primary cause since neuronal responses are seen in-vitro even with negligible change in temperature (Yoo et al. 2022), and human experiments are generally performed at low intensities that produce minimal heat (Lee et al. 2022). There are several theories that involve the direct interaction of the pressure field with the plasma membrane, which can have a number of effects. For example, stress in the membrane gets transferred to ion channels, causing their conformational state to change (Jerusalem et al. 2019). Initial work showed that voltage-gated sodium and calcium channels can respond directly to sonication, even though these channels are not categorized as mechanosensitive (Tyler et al., 2008). Subsequent work has largely focused on mechanosensitive channels (e.g., TRPP1/2, Piezo1), which can respond robustly to sonication. Such studies often observe that voltage-gated channels (non-mechanosensitive) are activated only as a secondary, downstream effect (Sorum et al., 2021; Yoo et al, 2022). Specifically, the mechanosensitive channels open in response to sonication, and this causes ion movement that produces changes in transmembrane voltage, which in turn modulates the voltage-gated channels. In addition to neurons, glial cells such as astrocytes have been shown to exhibit calcium influx via TRPA1 channels, which can in turn influence neuronal function through glutamate release onto NMDA channels (Oh et al., 2019; Jin et al., 2024). An alternative mechanism proposes that intramembrane cavitation, or the formation of nanoscale gas voids within the lipid bilayer, could contribute to the ultrasound response. This model has been used to explain the polarity-dependent neuromodulator responses (i.e., excitation versus suppression) seen in animals and humans (Plaksin, Kimmel, and Shoham 2016). However, there is also contradictory evidence that suggests intramembrane cavitation is not the only factor at play (Dell'Italia et al. 2022). Taken together, the evidence suggests that there could be multiple overlapping biophysical mechanisms that shape the neural response to tFUS.

An alternative strategy to understanding the neuronal response to ultrasound could be to focus on the axon alone, thereby avoiding any complex effects relating to synaptic transmission. Indeed, a model of white matter sonication has been developed (Felix et al. 2022). In this model, ultrasound-induced mechanical deformation of axons, particularly bending, transiently alters membrane polarization via the redistribution of bound charge within the lipid membrane. These changes are hypothesized to influence voltage-gated ion channel behavior by modifying local membrane potential. However, despite the potential utility of targeting the white matter, it has not yet been shown experimentally that white matter can be targeted for neuromodulation in humans. White matter sonication could be useful in that it could minimize the plasticity changes that are specific stimulus-induced calcium currents, and therefore this remains an important area for future testing. As a point of reference, sonication of axons in the peripheral nervous system can influence firing rate, although there is considerable variability in results between species, and some of these effects may not be present at intensities safe for human use (Blackmore et al. 2019, Yoo et al., 2017).

Defining the temporal response properties of neurons to tFUS stimulation requires some digression. Typically, the temporal response of a system is measured with a frequency response using sinusoidal inputs. However, the sonication frequencies used in tFUS (> 200 kHz) exceed the response time of neurons, including the voltage-gated sodium channels that underlie action potentials and the voltage-gated calcium channels that underlie synaptic release, both of which open and close on the millisecond timescale (McIntyre et al. 2004; Freeman et al. 2010, 2011). This is fundamentally different from electrical stimulation of neurons, in which stimulation can be applied transcranially at relatively low frequencies where brain oscillations occur (e.g., 10 Hz) (Cabral-Calderin and Wilke 2020). However, with electrical stimulation, sinusoidal waveforms are generally avoided in order to prevent electrochemical degradation of the electrode. Instead, the temporal response properties to electrical stimulation are often characterized by applying short-duration biphasic pulses at various rates (e.g., 5 Hz versus 50 Hz) (Freeman, Rizzo, and Fried 2011).

tFUS could take a similar approach as pulsed electrical stimulation by applying repetitive bursts of sonication and estimating the neural response to varying rates at which the bursts are delivered. However, such measurements are rarely reported, and this is likely due to the fact that tFUS is generally not aimed towards reproducing dynamically varying firing rates. As a point of comparison, cochlear implants use electrical stimulation of the auditory nerve with the goal of replicating spike trains that include rapid changes in firing rate that encode dynamic speech patterns. This is fundamentally different than deep brain stimulation (DBS) for Parkinson's Disease or Spinal Cord Stimulation (SCS) for chronic pain, where the goal is not to replicate a dynamically changing firing rate of neurons, but rather to alter the overall state of the neural network to have some desired clinical effect. In that sense, tFUS is more analogous to DBS and SCS than it is to cochlear implants or retinal implants. Therefore, most studies that examine the neural response properties to tFUS do not report temporal resolution, although some recent efforts are aiming to deliver tFUS at rates that entrain brain oscillations using the so-called theta-burst protocol (Yaakub et al. 2023; Kim et al. 2024; Grippe et al. 2024).

Despite the fact that tFUS is not usually used to replicate rapid changes in firing rate (e.g., as with cochlear implants), the temporal pattern of sonication pulses can still have a significant impact on the neural response. In this case, a "pulse" is a single burst of sonication at the carrier frequency (e.g., 1 ms duration pulse of a 500 kHz sinusoid). There have been extensive efforts to categorize the neural response across parameter space, including as a function of pulse duration, pulse repetition rate, duty cycle, intensity, carrier frequency, and the overall duration of sonication (Dell'Italia et al. 2022; Nandi et al. 2024; Zhang et al. 2021; Darmani et al. 2022). Such efforts aim to classify a given range of stimulus parameters as having either an excitatory or inhibitory effect (e.g., see Figure 4 of Dell'Italia et al., 2022), but despite some trends, there remains too much variability between studies to make any definitive conclusions. The field will benefit from future studies that vary stimulation parameters within a given session in order to keep as many variables as constant as possible (Fomenko et al. 2020).

Another important lesson that tFUS research can learn from decades of work on electrical stimulation is that the neuronal response is dependent not only on the direct interaction of the applied stimulus with individual cells, but also on downstream effects and network level effects. These include extracellular potassium accumulation, recurrent excitatory and inhibitory dynamics, and importantly, changes in synaptic strength due to long-term potentiation (LTP) and depression (LTD) (Halnes et al. 2016; Cooke and Bliss 2006). These processes operate over wide range of timescales, from the sub-millisecond dynamics of voltage-gated ion channels, to the persistent effects of plastic changes to the synapse, which can last for days or weeks. Given the observation that tFUS can produce an influx of calcium (Tyler et al. 2008) - a key mediator in activity dependent synaptic plasticity - it is not surprising that effects are observed the extend beyond the sonication period. Indeed, both animal and human studies have demonstrated long-lasting neurophysiological changes following tFUS of primary somatosensory cortex, with altered evoked potentials persisting for 35 minutes in anesthetized rats (Yoo et al., 2018), and functional connectivity changes lasting at least an hour in humans (Kim et al., 2023). Interestingly, these studies both observed the long-lasting effect even though the sonication parameters were very different, with 70% duty cycle in humans and 5% duty cycle in rats. Understanding how sonication parameters interact with region-specific cell-types remains a key challenge to developing predictive models of tFUS effects. . Finally, even when the single neuron response to tFUS is fully characterized, it will be necessary to take into account the complex feedback and feedforward connections of excitatory and inhibitory pathways (Nandi et al. 2024).

A closely related consideration is the temporal profile of these effects, and in particular, how long the effects persist after the sonication period has ended. While some studies focus only on the effects that occur during the sonication itself ("online effects"), other studies explore the persistent effects that last beyond the sonication period ("offline effects"). These offline effects are particularly relevant for clinical applications, such as chronic pain or depression, where the goal is to produce long-lasting relief. Systematic studies evaluating the impact of sonication parameters on the duration of offline effects are in the early stages (see review from (Bault, Yaakub, and Fouragnan 2024)). However, accumulating evidence suggests that even short periods of sonication (e.g., tens of seconds) can yield long-lasting modulation of cortical and subcortical function. Animal studies in rats, pigs, and non-human primates have demonstrated effects that last minutes to hours following in-vivo stimulation (Darmani et al. 2022). For example, in one primate study, offline effects last > 1 hour after 40 seconds of sonication of both cortical and subcortical structures (Verhagen et al. 2019). Interestingly, using resting state fMRI, they were able to show that tFUS produced changes in functional connectivity to other brain areas, and this effect was remarkably site specific in that it was not present when sonication was applied to a control region that was millimeters away from the target. In humans, similar offline effects have been observed: for example, sonication of prefrontal cortex produced changes in functional connectivity and associated mood changes at 20 minutes after sonication (Sanguinetti et al. 2020). Tracking the duration and evolution of such effects in humans is logistically challenging, and outcomes could vary depending on the method used for monitoring (e.g., fMRI and EEG versus behavioral reports).

To conclude the discussion on the neural response to tFUS, we provide three more observations that may be helpful for those new to the technology. First, sonication of sensory or motor cortex in humans does not necessarily produce a sensory percept or overt muscle contraction as with animal studies, where pressure levels can be higher. For example, when sonicating primary visual cortex in humans, a recent study showed no visual percepts were elicited (Nandi et al. 2023), despite an earlier study for which visual percepts were observed (Lee et al. 2016). As a point of comparison, electrical stimulation of visual cortex reliably produces visual percepts, and therefore, it could be that low-intensity tFUS stimulation does not elicit the same transient bursts in spiking that are produced with electrical stimulation. More studies are needed to better understand whether tFUS can be tuned to reliably produce sensory percepts. Second, there have been pilot studies that have shown that tFUS can produce arousal of patients in a minimally conscious state (Cain et al. 2021). This illustrates the potential of tFUS to fundamentally alter brain state in a way that is distinct from other offline effects, whose impact can wear off in hours. This line of research, while primarily aimed at clinical treatment for disorders of consciousness, has great potential for contributing to our understanding of the neural circuits underlying general arousal. Third, it is important to note that sonication may generate audible sounds that elicit responses in auditory pathways, potentially causing confounding effects in higher-level brain areas. Various mitigation strategies are employed, such as the use of masking sounds or a modulation envelope on the sonication waveform (Pellow, Pichardo, and Pike 2024), or by using an unfocused sonication as a control condition (Fan et al. 2024).

*Commercial Technology for tFUS and Brain Monitoring*

One of the key elements of the tFUS system is the piezoelectric transducer that is used to generate pressure waves that are applied to the head. The transducer can be directly coupled to the scalp with gel, or it can be coupled through a water-filled bladder or compressive hydrogel. A key function of the transducer is to focus the pressure beam to a specific target region in the brain. Such focusing can be achieved either with a curved transducer, or through beamforming with a series of flat elements. For example, a curved transducer can be spherical in shape, where a larger radius of curvature transducer can be used to focus on deeper targets. Conversely, beamforming can be achieved with a phased array of transducer elements, allowing steering of the beam in 3D, or an annular ring design, which allows steering in depth only (Chaplin, Phipps, and Caskey 2018). The electronics that drive the transducer are designed to deliver high-power through an impedance matching circuit (Javid, Ilham, and Kiani 2023). Most often, these electronics are not designed custom, but rather they are purchased as part of a full ultrasound system, or they are procured as benchtop power amplifiers.

The skull exhibits significantly more attenuation of the pressure wave than soft tissue (~20 dB/cm/MHz versus ~0.5 dB/cm/MHz) (Darmani et al. 2022). Because skull thickness can vary significantly across subjects, a CT scan is often performed so that the intensity of the sonication can be adjusted to achieve a desired pressure at the focal region. An alternative approach to tailoring the sonication intensity for each subject is to use the same sonication intensity across all individuals in a given study, regardless of skull thickness. The downside to this approach is that the actual peak pressure at the target location may vary across individuals, but the upside is that there is much less risk in accidently exposing the subject to pressure levels that exceed regulatory guidance (Lee et al. 2016). With this strategy, it is still useful to have a CT to inform the data analysis because one can incorporate skull thickness into numerical stimulations, allowing the actual peak pressure to be estimated for each individual subject. For example, if there is an effect that is highly dependent on peak intensity of the pressure field, then one might observe a trend across subject.

If skull thickness varies considerably over the region under the transducer, the resulting pressure beam can become defocused. Such defocusing can be corrected for with phased arrays, where the pressure beam can be electronically steered. Conversely, spherical transducers cannot correct for such defocusing, although the transducer can be repositioned to provide some partial correction to ensure the peak pressure is on the desired target. To correct for defocusing with spherical transducers, there is research aimed at using a custom 3D-printed acoustic lens that goes in between the transducer and the scalp (Maimbourg et al. 2020; Jiménez-Gambín et al. 2019). A given acoustic lens would be tailored to a particular subject, providing a cost-effective means of defocus correction. In all of these systems, including spherical transducers and phased arrays, intensive numerical simulations with the CT and MRI images are required. In some cases, laboratories have developed custom simulation tools (Shin et al. 2024), and in other cases, the simulations are performed as part of the built-in functionality of commercially-available tFUS systems.

We will provide some examples of tFUS systems that are commercially available, but this list is not comprehensive, as there are many companies emerging in this field. An example of an end-to-end complete tFUS system is the ExaBlate from InsighTech, offering a 1,024-element phased array. This system was originally designed for high-intensity ultrasound for thermal ablation, and is being repurposed for low-intensity applications, including neuromodulation, but must be used within the MRI bore (Mahoney et al. 2023). Another phased-array system is from the company NaviFUS, offering 256-elements (Lee et al. 2022). The company Brainbox offers the NeuroFUS system with an annular ring design that can adjust the pressure beam in depth only and not in 3D (Yaakub et al. 2023). The company Spire has developed a unique system consisting of two spherically-shaped grids with 126 elements that are placed on opposite sides of the head (Riis et al. 2024). Instead of using a CT for correcting for skull defocusing, this system applies and measures ultrasonic signals to infer the beam pattern. Another unique system is from the company Attune, consisting of a head-worn device with two 64-element phased arrays positioned on either side of the head (Fan et al. 2024). Finally, Openwater is offering a 64-element phased array that is suitable for research laboratories, allowing users to program the beamforming directly (Bawiec et al. 2025). Also, the Openwater system allows multiple arrays to be daisy chained, which can improve targeting of complex anatomical structures.

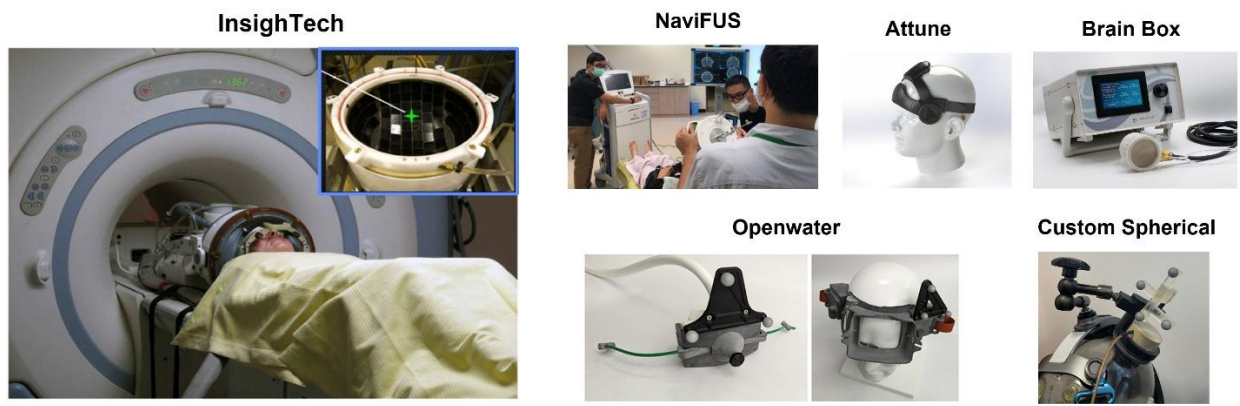

Figure 1. Images of various transcranial ultrasound systems, including phased-arrays (InsighTech, NaviFUS, Attune, Openwater), annular transducers (Brainbox), and spherical transducers (custom, see Lee et al., 2015).

The method for fixing the head relative to the transducer varies across these systems. For example, the InsighTech system uses a stereotactic frame to hold the head in place, while the Attune system is naturally fixed to the head through a custom-fitted cap. Depending on the size of the aperture and the distance between the transducer and the scalp, some systems are better optimized for the deep brain as opposed to cortical targets. Head shaving is usually not required, except for in some specific clinical studies (Mahoney et al. 2023). Although all of these systems have been used in humans, not all systems are approved for use in healthy human subjects, which is subject to approval of the company as well as from the relevant government agencies (e.g., FDA) and local institutional review board (IRB).

There are a variety of brain monitoring techniques that can be used to evaluate the effects of tFUS in humans. One of the most common methods is EEG, which measures electrical activity of the brain using scalp electrodes. EEG can be used, for example, to monitor the impact of sonication on various brain rhythms (Legon et al. 2018). Alternatively, EEG can be used to measure the impact of sonication on an evoked response, such as those measured in primary somatosensory or visual cortex (Lee et al. 2015, 2016). Another common method for brain monitoring is fMRI, which measures activity-dependent changes oxy/deoxy hemoglobin content in cerebral blood. Although fMRI does not capture rapidly-varying electrical activity, it exhibits high-spatial resolution across the entire brain, providing a powerful method for evaluating the specific brain networks engaged during tFUS of a particular structure (Nakajima et al. 2022). Additionally, resting state fMRI (rsfMRI) can be used to evaluate changes in functional connectivity between brain areas following tFUS. For example, sonication of anterior and posterior cingulate cortex show changes in functional connectivity across the default mode network and salience network (Yaakub et al. 2023). Other techniques that have been used, although less commonly, include magnetic resonance spectroscopy (MSR) for measuring neurotransmitter levels (Bault, Yaakub, and Fouragnan 2024), and transcranial magnetic stimulation, which does not measure neural activity, but can be used to produce an evoked response that can be modified with tFUS (Fomenko et al. 2020).

**Existing Tools for Studying Conscious Perception - Correlational and Causal Evidence**

There are many techniques available for monitoring and manipulating brain activity to study consciousness. The following sections provide a concise overview of several key approaches that have contributed to our current understanding of conscious perception.

*Techniques for Monitoring Brain Activity in Consciousness Research*

The most common methods for monitoring brain activity for investigating conscious perception are electroencephalography (EEG), functional MRI (fMRI), and intracranial recordings. While these tools have greatly improved our understanding of the neural correlates of consciousness (NCC), they remain limited in their ability to identify the underlying mechanisms that give rise to conscious experience (Dehaene and Changeux, 2011; Michel, 2019).

EEG offers high temporal resolution and has revealed relationships between brain rhythms (e.g., alpha, beta, gamma) and perceptual awareness. However, its poor spatial resolution and vulnerability to scalp and volume conduction make it difficult to localize the source of neural activity with precision. Moreover, much of the EEG literature is correlational, and it remains unclear whether the observed oscillatory patterns reflect the NCC itself, modulations in attention, or downstream consequences of perception. Aside from large-scale brain oscillations, EEG can also be used to record event-related potentials (ERP) in response to a given task. For example, ERPs such as P3b or perceptual awareness negativity (PAN) show some sensitivity to awareness, but report-related confounds can make interpretation difficult (Cohen, Ortego, and Kyroudis 2020). In addition, there are differences across sensory modalities, where PAN is clearer in somatosensory experiments than visual or auditory tasks, challenging the idea that a specific EEG feature serves as a universal NCC (Dykstra et al., 2017).

Another commonly used brain monitoring technique in consciousness research is fMRI, which offers excellent spatial resolution and whole-brain coverage. However, fMRI suffers from poor temporal resolution because the underlying signal reflects the hemodynamic response rather than sensing electrical activity directly. Also, it can be difficult to interpret signals from certain brain regions where neurons exhibit mixed selectivity, meaning they respond to a wide range of task-related variables (Malach, 2021). This is especially true in frontal areas of the brain, where this overlapping activity leads to statistical ambiguity, making it difficult to determine if the observed signals reflect conscious perception versus other cognitive processes, such as attentional control and decision making (Kronemer et al., 2022).

Neurosurgical techniques performed on awake patients offer a unique opportunity to explore the neural basis of conscious perception. During these procedures, researchers can use electrocorticography (ECoG) and intracranial depth electrodes to record high-resolution electrical activity from the cortex or deep brain structures. However, such recordings are limited to clinical evaluation, usually with small brain coverage.

Collectively, these monitoring tools offer correlational insights into brain activity associated with perception, but none provide the spatiotemporal precision and causal specificity required to definitively identify the neural circuits and mechanisms that generate conscious experience.

*Techniques for Brain Stimulation in Consciousness Research*

While monitoring tools offer correlational evidence, brain stimulation techniques such as transcranial magnetic stimulation (TMS), transcranial direct and alternating current stimulation (tDCS, tACS), and direct electrical stimulation of the brain during surgery can be used to provide causal insights into the neural basis of consciousness. However, these methods also have limitations for isolating the NCC.

TMS allows for noninvasive, targeted disruption or excitation of cortical areas. When applied to early sensory areas like V1, TMS can suppress awareness or induce phosphenes, offering evidence of involvement in conscious perception (Bachmann, 2018). However, the neural response likely involves indirect modulation of downstream brain regions, making it difficult to isolate the specific neurons involved causally in generating perception (Bareham et al., 2021). The two primary downsides of TMS are the coarse spatial resolution and the inability to target deep brain structures. As far as notable experimental findings, TMS has been shown to dissociate awareness from performance, inducing a blindsight-like effect (see examples in Odegaard et al., 2017). However, these results are under debate because residual feedback or unconscious processing could explain such dissociations (Bareham et al., 2021). Another example is the use of TMS on prefrontal areas, providing evidence for modulating conscious awareness (Michel, 2022). But the role of the prefrontal cortex in awareness remains under debate, partly because unilateral stimulation may be insufficient, and also because frontal regions often participate in post-perceptual decision making (Kozuch, 2024).

Another neuromodulation technique involves the direct electrical stimulation of the brain during neurosurgery (Raccah, Block, and Fox 2021). Such stimulation has been invaluable for mapping out cortical functional maps, and humans report a range of region-dependent perceptions, including phosphenes, auditory and tactile perception, as well as more complex sensations, such as fear, when stimulating the prefrontal cortex. However, as with TMS, local stimulation can propagate through complex interconnected networks, making it difficult to identify the true neural substrate of conscious perception. A classic example is the stimulation of primary visual cortex (V1), where it is not clear whether the resulting visual percept is due to activity within V1, or whether the percept is the result of activity in downstream areas (Pollen, 2004).

To illustrate this point more broadly, consider a simplified hypothetical example that involves the search for the neural substrate of pain (Figure 2). In this example, we will assume that the sense of pain is causally produced by one of the following: (1) activity in Synapse #1, located in anterior cingulate cortex (ACC), (2) Synapse #2, located in periaqueductal gray (PAG), or (3) both Synapse #1 and #2 must be active for the conscious perception of pain. If TMS or tDCS is applied to ACC, then this will directly influence Synapse #1, but it will also indirectly influence Synapse #2 due to their synaptic connection, making it difficult to interpret experimental results. In other words, the immense amount of network connections between brain areas will mean that stimulation of any one structure will produce a cascading effect of feedforward and feedback activity, complicating the interpretation for causality between neural activity and the sensation of pain. Ideally, experimental techniques can address this challenge using one of several methods: eliciting excitatory versus suppressive effects, by timing differences between the stimulus onset and the percept, or by selective targeting of the white matter that enters or leaves a given region. Importantly, TMS and tDCS both have limitations; neither can target PAG because of its deep location in the brain, neither is capable of eliciting excitatory versus suppressive responses, and neither has the spatial resolution capable of targeting white matter tracts that connect these structures. These limitations highlight the need for alternative approaches, such as tFUS, which offers the potential for greater spatial specificity, deeper penetration, and the capacity to produce excitatory versus suppressive effects by tuning sonication parameters.

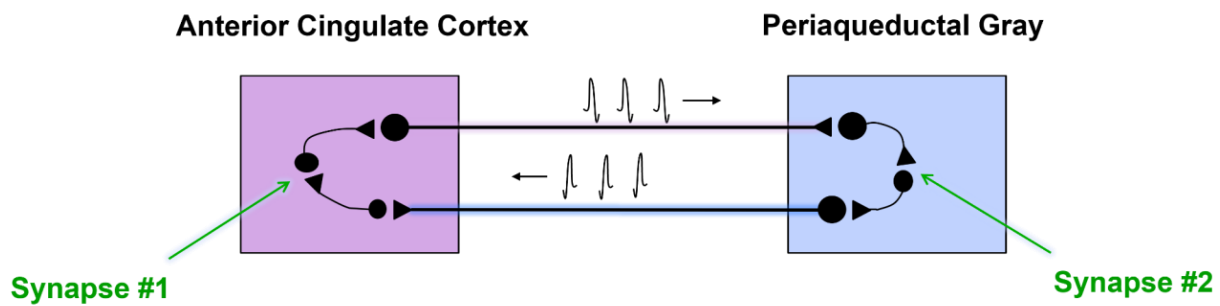

Figure 2. Simplified schematic of a hypothetical neural circuit in which pain perception arises from synaptic activity at either or both of two candidate sites: Synapse #1 and/or Synapse #2. The two brain regions, anterior cingulate cortex (ACC) and periaqueductal gray (PAG), are connected with bidirectional signaling with action potentials transmitted in both directions. As a result of this connectivity, experimental perturbation of either brain region will influence activity in both synapses, complicating efforts to localize the neural substrate responsible for the perception of pain.

# A Roadmap for Using tFUS to Investigate the Neural Substrate of Conscious Perception

*Contemporary Theories and Debates in Consciousness Research*

Theories on the neural basis of consciousness vary widely, with some specifying particular brain structures and neural pathways critical for consciousness, while others focus on defining general features of neural activity required for conscious experience, without committing to specific anatomical structures. These latter theories often emphasize the functional characteristics of neural interactions, such as the integration of information or the nature of feedback versus feedforward connections, leaving open the identification of specific neural substrates involved. It is important to note that for any particular theory to be successful, it must eventually specify the neural substrate required, even if early versions of the theory are agnostic about the anatomy. Since these theories have been described in detail elsewhere (Seth and Bayne 2022; Mudrik et al. 2025), we will summarize some of the main theories only briefly here, with an emphasis on the position each theory takes on the neural substrate required to produce consciousness.

1. The **Global Workspace Theory (GWT)** suggests that consciousness arises when information becomes "globally available" to various cognitive systems within the brain (Stanislas Dehaene and Changeux 2011). This theory emphasizes the importance of distributed networks, primarily involving the frontal and parietal regions, which facilitates extensive communication across neural systems. At the heart of GWT is the concept of "ignition," a rapid surge of neural activity that spreads through this network, signifying that information has reached a sufficient level of accessibility. This widespread accessibility enables the engagement of multiple cognitive functions such as perception, attention, memory, decision-making, and language processing. GWT tends to focus on the connectivity patterns that support integrative cognitive activities rather than on specific brain structures themselves. Subcortical structures are considered to play a supporting role (e.g., arousal).

2. **Integrated Information Theory (IIT)** posits that consciousness correlates with the level of integrated information within a network, quantified as phi (Φ) (Tononi 2012). IIT particularly emphasizes the cortex due to its extensive network of interconnections, including both feedforward and feedback, allowing it to achieve high levels of Φ. Feedback connections, in particular, enhance the integration of information by creating dynamic loops within neural circuits, potentially increasing Φ. While IIT predominantly highlights the cortex, it does not exclude subcortical areas from potentially contributing to consciousness. Any part of the neural network, including subcortical structures, could contribute to consciousness if it is part of a network that exhibits substantial integrated information. There has been recent scrutiny of this theory; see (IIT-Concerned et al. 2025).

3. **Local Recurrence Theory (LRT)** posits that recurrent neural connections, rather than purely feedforward connections, are fundamental to the emergence of consciousness (Lamme 2006). This theory emphasizes the significance of feedback loops from high to low-level sensory areas, suggesting that such interactions enable sensory information to reach conscious awareness by enhancing and sustaining neural activity within these regions. The visual cortex is often used as a primary example due to its well-documented architecture of recurrent connections, but the theory applies generally to other sensory cortices. This theory does not explicitly state whether activity in early sensory cortex is necessary and sufficient for conscious awareness, or whether activity in higher-order association areas is also required. Also, while local feedback within cortex is the primary emphasis of LRT, this theory does not rule out the possibility that thalamo-cortical feedback may play a supporting role as well.

4. **Higher Order Theories (HOTs)** are a group of theories positing that consciousness emerges from a form of awareness of one's mental states. These theories thus link consciousness to metacognition—the capacity to monitor and control one's mental states (Lau and Rosenthal 2011). These theories emphasize the importance of cortical structures, particularly the prefrontal cortex and associated higher associative areas, which are known for their role in complex cognitive functions like metacognition, planning, and decision-making. While the primary focus of HOTs is on cortical involvement, these theories generally regard subcortical structures as playing a supportive, albeit indirect, role. For example, regions like the thalamus and limbic system are crucial for regulating arousal and emotional states that underpin the wakefulness and alertness necessary for conscious experience. However, in the framework of HOTs, these areas are seen more in terms of facilitating or modulating the cortical activities that are directly involved in the reflective processes that characterize conscious thought.

This list is far from exhaustive (for other views, see, e.g., (Graziano and Webb 2015; Malach 2021; Merker 2007). Painting with broad strokes, recent debates about consciousness have focused on two competing theoretical views about perceptual consciousness and its functions. *Cognitivist* views propose that consciousness plays a role in supporting cognition by making perceptual information accessible to a broad set of cognitive mechanisms such as working memory, reasoning, and verbal report (as in global workspace theory; Dehaene, 2014), by enabling metacognitive monitoring of these percepts (as in higher order theories; Lau, 2022), or by enabling the control of attention (Graziano and Webb 2015). In contrast, *non-cognitivist* views argue that consciousness plays a functional role for perception itself, such as enabling perceptual processes like figure ground segregation or stabilizing representations over time (Lamme, 2015; Broday-Dvir et al., 2023). For example, one specific form of this view holds that conscious experience is constituted entirely by structured patterns of sensation, including basic sensory modalities (e.g., vision, audition, touch), interoceptive and affective states, and possibly more abstract sensory like qualities (e.g., emotional tones or feelings of familiarity). In this view, consciousness does not depend on information being made available for cognitive use or report. Rather, whenever a particular neural population is active with some particular features, a corresponding conscious sensation arises. Memory, language, and reasoning may access or be shaped by this stream of sensation, but they are not constitutive of it (e.g., Zeki 2008). These contrasting perspectives carry implications for which brain regions generate consciousness. According to global workspace theory, for example, consciousness depends on widespread broadcasting across frontoparietal networks that support cognitive access. In contrast, non-cognitivist and sensation-based theories typically emphasize localized activity in posterior cortical areas or subcortical structures as the primary substrates of consciousness (Lamme, 2006, 2018; Merker, 2013).

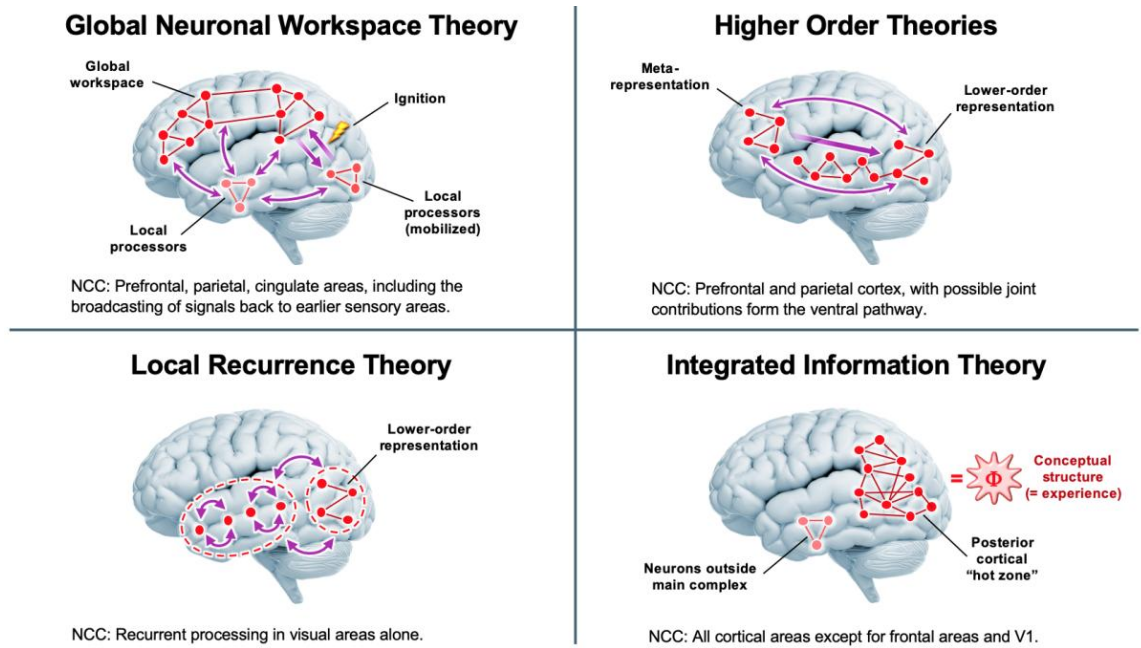

*Figure 3*. Neural substrates underlying conscious visual perception are illustrated for four major theories of consciousness. Each panel highlights specific brain regions implicated in the conscious processing of visual stimuli. NCC = Neural Correlates of Consciousness

A central empirical question in consciousness science concerns the role of the prefrontal cortex (PFC). Cognitivist theories, such as global workspace theory and higher order theories, predict that conscious perception depends on activity in prefrontal and parietal regions, which allow perceptual contents to be accessed, maintained, monitored, and reported. In contrast, non-cognitivist theories typically argue that consciousness arises within perceptual systems themselves. In this latter view, prefrontal activity may well be the source of conscious contents, but it is not the site where the content becomes conscious (Merker, 2012).

Disentangling these views empirically has proven difficult, in part because many experimental paradigms confound conscious perception with cognitive processes associated with reporting it. This issue is commonly referred to as the report confound and it is particularly important in studies where participants are required to give explicit ratings of visibility (Sandberg and Overgaard, 2015; Overgaard and Sandberg, 2021) or confidence (Michel, 2023). Critics have argued that PFC activation observed in such studies may reflect the process of reporting consciousness, rather than consciousness itself (Tsuchiya et al., 2015).

In response to these experimental challenges, researchers have developed no-report paradigms, which aim to infer conscious states without requiring explicit reports, for example, by tracking involuntary eye movements like optokinetic nystagmus during binocular rivalry (Frässle et al., 2014). Crucially, many of these no-report studies still find reliable PFC involvement in conscious perception when using sensitive enough measurement techniques (Kapoor et al., 2022; Michel and Morales, 2019; Michel, 2022). These findings support the cognitivist view that PFC activity is part of the neural substrate of consciousness, even in the absence of overt reporting (although see Block 2019, 2020).

Another challenge for identifying the neural substrate of conscious perception comes from uncontrolled differences in unconscious processing between the 'conscious' and 'unconscious' experimental conditions. Conscious experience is preceded by a cascade of unconscious processes. Unconscious processes also run in the background of every conscious experience. Experimental manipulations used to render stimuli unconscious do not simply turn consciousness off (Breitmeyer 2015); they also affect unconscious processing that occurs before, and runs parallel to, conscious processing. It follows that the contrast between the 'conscious' and 'unconscious' conditions likely significantly overestimates the activity associated with conscious processing. To illustrate with an extreme example: one way of suppressing visual consciousness of a stimulus would be to ask participants to close their eyes. In this hypothetical experiment the contrast between full consciousness of the stimulus and the eyes-closed condition would not reveal the neural correlates of consciousness. Indeed, consciousness is not the only difference between the two conditions. The whole chain of unconscious perceptual processing that precedes consciousness, and runs alongside it, differs as well. While our current situation with existing experimental tools is not as extreme, most methods for suppressing consciousness might work a bit like this, in that they interrupt perceptual processing quite early (though some interrupt perceptual processing later than others, see (Chakravarthi and Cavanagh 2009). As a result, differences resulting from asymmetries in (unconscious) perceptual processing between the conscious and unconscious conditions constitute a significant confounding factor when studying the neural correlates of consciousness.

To alleviate this concern, one needs to keep the conscious and unconscious conditions as similar as possible in all respects, while still suppressing consciousness. This implies preserving as much unconscious processing as possible in the unconscious condition—thereby suppressing perceptual processing as close to the finish line of consciousness as possible. Using this methodology, the resulting contrast between conscious and unconscious conditions is more likely to reflect activity that is specific to consciousness, rather than differences in factors unrelated to consciousness (e.g., overall discriminability of the stimulus, or strength of sensory signals). Unfortunately, this problem has not attracted as much attention as the challenge of controlling for report-related activity (though see (Lau 2008), but it is no less important. We discuss some ways of reducing the impact of this problem below.

Overall, these methodological efforts are helping to clarify whether consciousness depends on access-related processes involving the PFC, as cognitivist theories suggest, or whether it can arise independently in perceptual circuits, as non-cognitivist and sensation-based views propose. As the field continues to refine experimental designs and neural measures, these competing models make divergent predictions about where in the brain consciousness occurs, and about what kinds of neural activity are necessary for a sensory state to be experienced rather than merely processed unconsciously.

*The Importance of Visual Psychophysics and Blindsight in the Study of Consciousness*

There are two ways of decreasing asymmetries in perceptual processing between conscious and unconscious conditions, which could get us closer to studying consciousness itself. The first strategy is to develop 'performance-matched' experiments in which discriminability is just as high in the conscious and unconscious conditions (Morales, Odegaard, and Maniscalco 2022). Using discriminability as a proxy for the strength of perceptual processing, this allows one to make sure that perceptual processing is maintained as similar as possible between the conscious and unconscious conditions. To do so, one could leverage dissociations between discriminability and consciousness observed in multiple phenomena, such as: the perception of ensemble statistics (Elosegi, Mei, and Soto 2024; Sekimoto and Motoyoshi 2022), in the phenomenon of lag-1 sparing in the attentional blink (Pincham, Bowman, and Szucs 2016; Recht, Mamassian, and de Gardelle 2019), in a phenomenon called 'subjective inflation' (Rahnev et al. 2011; Knotts et al. 2019; Odegaard et al. 2018), and in aphantasia (Keogh and Pearson 2018; Pounder et al. 2022; Weber et al. 2024). For example, Lau & Passingham (2006) conducted a task where participants had to judge (1) whether a briefly presented stimulus was square or diamond, and (2) report whether the stimulus was seen or unseen. Using metacontrast masking, they manipulated the stimulus onset asynchrony between the stimulus and the mask to create two conditions that resulted in near-equal discrimination performance, but different reports of how frequently the stimulus was seen. They found that the only area associated with variations in consciousness in this paradigm is the dorso-lateral prefrontal cortex (dlPFC). The main drawback of performance-matched paradigms is that they generally require tedious psychophysics and that the effect sizes are generally small—on the order of differences of a few percentage points in reported visibility between the conditions.

The other strategy for studying variations in consciousness while maintaining discrimination capacities relatively equal between conscious and unconscious conditions is the phenomenon of blindsight (Weiskrantz 2009). Blindsight refers to preserved visual capacities in the absence of reported visual experience, following lesions to the primary visual cortex (V1). A wide variety of visual capacities are preserved in blindsight, including movement and shape discrimination, attention to stimuli in the 'blind' field, and automatic categorization of facial emotional expressions (Weiskrantz, 2009). Blindsight is invaluable for consciousness research because it is currently the only phenomenon in which one can reliably study the contrast between conscious and unconscious processing while keeping visual capacities relatively well preserved across conditions.

For these reasons, developing a technique for reliably inducing blindsight-like behavior in healthy subjects can be regarded as a major milestone in consciousness research. There are two main signatures of blindsight that would indicate blindsight-like behavior in healthy subjects. First, blindsight is characterized by a dissociation between 2-intervals forced-choice detection (2IFC)—where a target can be presented in one of two temporal intervals and the subject is forced to say which one—and Yes/No detection—in which a single interval may or may not contain the target and the patient has a choice between saying that the target was present or absent. The framework of Signal Detection Theory (Macmillan and Creelman 2005) predicts that performance on those two tasks, measured by an index called *d'*, should be such that 2IFC *d'* is superior to Yes/No *d'* by a factor of $\sqrt{2}$. But this mathematical relation breaks down in blindsight: Yes/No detection remains much poorer than 2IFC detection even once corrected by a factor of √2 (Azzopardi and Cowey 1997). The second signature of blindsight is the absence of metacognitive sensitivity to correct versus incorrect responses on a discrimination task. For instance, Persaud et al. (Persaud, McLeod, and Cowey 2007; Persaud et al. 2011) found that, in a discrimination task in which his performance was 75% correct, blindsight patient G.Y. was apparently incapable of betting on his discrimination responses appropriately. Namely, he did not appropriately abstain from betting when his responses were incorrect and bet when they were correct, thereby indicating that the patient had no idea whether his responses to visual cues were correct or incorrect.

To date, many studies have failed to induce similar effects in healthy subjects using the main suppression technique used in consciousness research, namely, visual masking (Balsdon and Clifford 2018; Peters and Lau 2015; though see Elosegi, Mei, and Soto 2024 and Amerio et al. 2024 for promising results). Perhaps the most promising case of blindsight-like behavior was obtained by (Rounis et al. 2010) who created conditions with matched discrimination performance and yet unequal metacognitive sensitivity after applying theta-burst transcranial magnetic stimulation to PFC. However, as explained above, relying on tFUS would clearly provide a superior method to investigate this question.

*Using tFUS to Evaluate the Role of Prefrontal Cortex in Conscious Perception*

For the reasons discussed above, the use of tFUS could solve two of the most challenging problems in the field at once. First, targeting of brain structures such as dlPFC provides a unique opportunity to resolve one of the most prominent debates in the field—elucidating the role of PFC in conscious perception. Second, the prospect of replicating the behavioral signatures of blindsight in healthy subjects—such as the dissociation between Yes/No and 2AFC detection, and metacognitive impairments, brings with it the promise of allowing subsequent research to study consciousness while maintaining perceptual processing as similar as possible between the conscious and unconscious conditions. Such a finding would mark a pivotal advancement in consciousness research by making it possible to reliably study the contrast between conscious and unconscious processing.

tFUS possesses a few characteristics that make it particularly well-suited to answer these two important problems in consciousness research. As mentioned in Section 2, regarding the question of PFC's role in consciousness, tFUS has the advantage of being able to apply bilateral stimulation, and application of bilateral tFUS to mPFC has been successfully implemented in paradigms which incorporate both excitatory and suppressive stimulation (Kim et al. 2022). Bilateral stimulation is critically important to conclusively answer whether PFC is involved in consciousness, as neuroplasticity following unilateral PFC damage has shown that frontal regions can often compensate for impairments to one hemisphere (Voytek et al. 2010). To evaluate the role of dlPFC or vlPFC in consciousness, the yes/no (yn) detection task and 2AFC task (i.e., technically speaking, a 2 interval-forced-choice task) from previous blindsight work (Azzopardi and Cowey 1997) could be adapted for normal observers, and a within-subject design that includes a *sham condition* and *suppressive tFUS stimulation condition* to these areas could be applied. If observers are asked to give perceptual judgments and confidence ratings after every trial, several hypotheses could be tested.

First, one could evaluate if dlPFC/vlPFC suppressive stimulation lowers the 2AFC/yn sensitivity ratio. Should suppression of PFC increase this ratio (as seen in blindsight patient G.Y.), it would support the conclusion that frontal areas may play a role in distinguishing seen from unseen information (Lau and Passingham 2006). Second, one could also analyze if dlPFC/vlPFC suppressive stimulation lowers confidence ratings in the 2AFC task. As lower metacognitive ratings have been noted as one of the defining features in blindsight (Ko and Lau 2012), any reduction in subjective ratings following suppression of PFC, whether they be confidence, visibility, or awareness ratings, would be noteworthy. Potentially, subjective ratings of more than one type could be tested to evaluate whether suppression of PFC affected all equally, or produced differential outcomes across confidence, visibility, and awareness. Third, one could evaluate whether suppression of dlPFC/vlPFC influences 2AFC meta-d' values. (Azzopardi and Cowey 2001) noted that the correspondence of confidence ratings with trial-by-trial accuracy was reduced in blindsight, potentially due to the jitter of internal "confidence criteria" used to evaluate perceptual evidence. Since PFC has been implicated in measures of metacognitive sensitivity (Fleming et al. 2010), and confidence ratings have been shown to have utility in studies of consciousness (Morales and Lau, n.d.), reduction of 2AFC meta-d' would provide further evidence of a role for PFC in awareness.

Importantly, tFUS could also be used to replicate and extend important performance-matched studies of awareness, such as Rounis et al. (2010). In that TMS experiment, participants performed a 2AFC task where they judged the spatial arrangement of two briefly-presented stimuli around fixation (square on left, diamond on right, or vice versa), and rated the visibility of the stimuli. Visibility was manipulated by altering the stimulus onset asynchrony (SOA) between the stimuli themselves and a metacontrast mask which followed their presentation. In the original work, theta-burst transcranial magnetic stimulation was applied to dlPFC, and the post-TMS results showed reduced visibility judgments (even with matched performance) and reduced metacognitive sensitivity. Concerns about the safety and replicability of this TMS procedure have been previously noted (Dehaene 2014); thus, tFUS presents a safer alternative to probe whether bilateral stimulation to dlPFC or vlPFC alters confidence, visibility, or metacognitive sensitivity in performance-matched designs. Questions still remain regarding how long tFUS can inhibit activity; the psychophysical tasks described here often take hundred (if not thousands) of trials (Azzopardi and Cowey 1997), and thus, it remains an open question of what the proper stimulation protocol may be to ensure that the effects of tFUS last throughout an entire experimental session. But recent work which provided stimulation at the start of each experiment block shows that reapplication during break periods can be an effective technique to target PFC activity (Jang et al. 2024). Thus, tFUS stands poised to answer important questions about the role of PFC on consciousness.

*Using tFUS to Evaluate the Role of Sensory and Association Cortices in Conscious Perception*

There are a number of areas of cortex that can produce clear sensory perception when stimulated electrically, during surgery, or with TMS. As discussed earlier, it remains unclear whether tFUS can produce similar effects in humans because of the differences in the neuronal response properties in the acoustic versus electrical domains, but there is ongoing work aimed at optimizing the sonication parameters to elicit percepts directly (e.g., visual, auditory, etc.). Assuming such percepts can be generated by tFUS, then one potentially useful set of experiments would be the use of multi-point stimulation, where one part of cortex is targeted to produce the percept, while a second brain region is suppressed. As with the TMS testing discussed above, the relative timing of the two tFUS stimuli can be varied to assess the role of feedback signals from higher areas to early sensory cortex.

Another hypothesis that can be evaluated with tFUS is the involvement of posterior association areas in generating conscious perception. These areas, encompassing parts of the parietal, occipital, and temporal lobes, are sometimes referred to as the "hot zone" because: (1) damage to these regions often leads to profound sensory deficits, and (2) direct electrical stimulation can result in perceptual effects across the visual, tactile, and auditory domains (Koch 2018). However, another possibility is that the neural substrate of conscious perception is not in the hot zone, but rather in downstream targets, either cortical or subcortical. A possible experiment using tFUS could involve selectively suppressing activity in the hot zone versus downstream targets, such as the frontal cortex or non-specific nuclei of the thalamus, while subjects perform a visual or tactile discrimination task. Such experiments would not only help clarify the role of the hot zone in perception, but would also allow a detailed map of the posterior association cortex by suppressing small, millimeter-scale regions. Before tFUS, such fine-scale mapping in humans was only possible during neurosurgery.

*Using tFUS to Evaluate the Role of Subcortical Structures in Conscious Perception*

Another major advantage of tFUS is that it can target subcortical structures, allowing researchers to evaluate the role of deep areas of the brain in conscious perception. While most major theories of consciousness emphasize the role of cortex, there are compelling arguments to consider the role of subcortical structures as well (Baron and Devor 2022; Redinbaugh and Saalmann 2024; Nieder 2021; Merker 2013; Fang et al. 2025). For example, one might assume that the neural mechanisms underlying perceptual sensations, such as vision or pain, began to emerge early in evolutionary history, even before the wide divergence of vertebrates (> 500 million years ago). This view follows a basic principle of evolutionary biology: it is more likely that any common features across vertebrates evolved only once in a common ancestor rather than evolving independently, multiple times, across a wide range of vertebrates (Striedter 2005). Considering the stark difference in anatomy between mammalian cortex and the pallium in birds and reptiles, this raises the question of whether older, subcortical structures might mediate conscious perception (i.e. phenomenal consciousness), perhaps in concert with cortex or pallium.

As discussed earlier, direct electrical stimulation of subcortical structures in humans has been performed during surgery for decades (Elias et al. 2021). Such studies can serve as a useful starting point to guide tFUS subcortical investigations. The number of potential subcortical targets is too large to cover here, but we consider a few example cases that illustrate the types of effects that can be evaluated. For example, direct electrical stimulation of the amygdala produces an array of sensations, including visual, auditory, olfactory, and somatosensory percepts, as well as emotional responses that included fear and anger, but also pleasure and joy, depending on the specific subnuclei being stimulated (Zhang et al., 2023). It remains unclear whether amygdala activity is necessary and sufficient to produce those sensations, but given the sparse sensory information available to the amygdala, it seems more likely that many of the sensations reported by the patients are due to neural activity in downstream targets that are driven by amygdala efferents. tFUS offers an excellent opportunity to localize the source of perception by non-invasively suppressing various targets that are downstream of the amygdala, for example, during a task that engages the amygdala (e.g., emotional facial recognition). In doing so, using tFUS could also help address the debate between the 'fear circuit' model (Fanselow and Pennington 2018), according to which amygdala activity is sufficient for the experience of fear, and the 'two systems model', according to which the amygdala is part of an unconscious 'defensive survival circuit' (LeDoux and Pine 2016).

tFUS can be used to improve our understanding of how the neural circuits underlying the conscious perception of sensory inputs (e.g., vision) are connected to higher-order areas involved in complex cognitive functions, such as attention and decision making. While the cortex is essential for such higher-order functions, there is evidence that subcortical structures are involved as well. For instance, it is already well established that certain subregions of the thalamus, such as pulvinar, play a critical role in visual attention (Kastner, Saalmann, and Schneider 2012; Kastner, Fiebelkorn, and Eradath 2020; Usrey and Kastner 2020; Saalmann and Kastner 2011). But recent work goes further in showing that specific subregions of the thalamus may directly contribute to distinct aspects of perceptual decision making; for example, results show that low-duty cycles shifted decision criteria towards a more conservative stance, whereas high duty cycle stimulation of the ventral anterior thalamus increased perceptual sensitivity (Jang et al. 2024). Additionally, previous work in nonhuman primates has shown that the pulvinar may uniquely contribute to the confidence in visual decisions, as pulvinar responses decreased when monkeys chose 'escape' options in a categorization task with an opt-out task, suggesting less confidence in their choices (Komura et al. 2013). These findings build on an increasing literature that has identified causal roles for subcortical structures in perceptual decision-making (Jun et al. 2021; Crapse, Lau, and Basso 2018).

Thus, we posit that tFUS can be used to further clarify the role of the thalamus (and specifically, pulvinar) in consciousness, in a manner that goes beyond recent demonstrations of its importance in disorders of consciousness (Monti 2021; Schnakers et al. 2018; Barra, Monti, and Thibaut 2022). Specifically, based on recent results implicating that the pulvinar contributes to confidence and the criterion while leaving sensitivity unchanged (Kaduk, Wilke, and Kagan 2024; Komura et al. 2013), we could take either the yn/2AFC paradigm from Azzopardi and Cowey (1997), or the matched-performance/difference-confidence paradigm (Morales, Odegaard, and Maniscalco 2022) from Rounis et al. (2010), and instead target the pulvinar nucleus of the thalamus. Our preliminary hypothesis would be that targeting the pulvinar should leave measures of perceptual performance (such as *d'*) unchanged, but would significantly alter confidence ratings and overall metacognitive sensitivity in our perceptual tasks. Potentially, tFUS to this region could be contrasted with suppression of other cortical areas (such as V1), which would presumably show changes in perceptual sensitivity across *sham* and *stimulation* conditions in a within-subjects design.

*Using tFUS for Investigating the Neural Substrate of Sensory Imagery and the Placebo Effect*

tFUS provides a powerful tool to further our understanding of the neural circuits engaged during sensory imagery, in which the subject imagines a particular sensory stimulus (e.g., picturing an object in your head, or imagining hearing a particular pitch or song). In the visual domain, recent work has challenged the view that imagery is essentially the reverse of visual perception in terms of the brain areas involved (Spagna et al. 2024). For example, while it is understood that V1 is essential for basic visual processes, its role in generating conscious visual imagery is still unclear. Also, the interaction of V1 with higher areas like V4 and the fusiform gyrus during complex imagery tasks, such as imagining faces or vivid scenes, is an active area of investigation. By applying tFUS to these areas during an imagery task, researchers could help understand whether V1 is sufficient for detailed imagery or whether integrative processes with higher visual areas are necessary (McNorgan 2012). Similar tests could be applied to other sensory modalities, such as the auditory domain, where possible targets include the primary auditory cortex, as well as association areas within the superior temporal gyrus and frontal cortex. Such tests could help elucidate the role of these structures in imagery tasks involving complex sounds or musical pitches (Greenspon, Pfordresher, and Halpern 2017).

Exploring mental imagery tasks that involve the integration of multi-sensory information can help understand how the brain combines information across different sensory modalities to form complex internal representations. For example, fMRI studies have shown that the superior temporal sulcus (STS) is involved in integrating visual and auditory information (Choi et al. 2023). However, it is unclear how the STS interacts with other brain regions during mental imagery, such as hippocampus or prefrontal cortex. Likewise, the posterior parietal cortex (PPC) has been implicated in spatial aspects of mental imagery, as well as motor planning, but there remain many open questions about the temporal dynamics of this signal integration, as well as the location of specific subregions of PPC involved in these tasks (Tian and Poeppel 2010). In both cases, either with the STS or PPC, there are opportunities for using tFUS to address these open questions by using spatially precise sonication of STS, PPC, and/or downstream or upstream targets, during an imagery task.

The placebo response in the context of pain treatments is another effect that can be explored with tFUS to help pinpoint the neural circuits that underlie the conscious perception of pain. The placebo effect is related to mental imagery in that both involve signals from higher-order brain areas that provide top-down modulation of the neural circuits involved in perception. To date, tFUS studies on pain have primarily focused on therapeutics rather than understanding the neural basis of pain. For example, tFUS has been evaluated for pain treatments by targeting areas such as the insula cortex (Legon and Strohman 2024) and anterior thalamus (Badran et al. 2022). There are many different potential nodes in the pain matrix that have yet to be explored (Kunar, 2020), including some areas that have been targeted with deep brain stimulation, such as ventral posterior thalamus and periaqueductal gray (Frizon et al. 2020). Interestingly, fMRI studies on pain have shown that the placebo effect involves brain areas related to affective and cognitive processes rather than directly altering nociceptive pathways (Botvinik-Nezer et al. 2024). This raises the possibility that higher-order brain areas can be targeted with tFUS to enhance or suppress the placebo effect to help localize the neural substrate of pain.

**Conclusion**

Transcranial focused ultrasound (tFUS) offers an unprecedented opportunity to advance the science of consciousness by enabling noninvasive, spatially precise, and depth-penetrating brain stimulation in humans. We reviewed key challenges in isolating the neural substrates of conscious perception and highlighted how tFUS can overcome the limitations of traditional stimulation techniques like TMS. The ability of tFUS to induce transient disruptions in brain activity makes it an ideal tool for experimentally dissociating conscious perception from unconscious processing and post-perceptual cognition. In addition, because tFUS can potentially be used to directly elicit or suppress perceptual experiences by stimulating multiple brain areas in a controlled manner, it offers a way to distinguish genuine substrates of perception from regions whose activity is correlative in nature. We propose a roadmap for using tFUS in conjunction with performance-matched and blindsight-like paradigms to identify the specific circuits and mechanisms underlying perceptual awareness. As tFUS technology continues to evolve, its integration into consciousness research promises to yield deeper insights into the brain mechanisms that give rise to our subjective experiences.

# References

Amerio, Pietro, Matthias Michel, Stephan Goerttler, Megan A. K. Peters, and Axel Cleeremans. 2024. "Unconscious Perception of Vernier Offsets." *Open Mind* 8 (June):739–65.

Azzopardi, P., and A. Cowey. 1997. "Is Blindsight like Normal, near-Threshold Vision?" *Proceedings of the National Academy of Sciences of the United States of America* 94 (25): 14190–94.

Azzopardi, P, 2001. "Why Is Blindsight Blind." *Out of Mind: Varieties of Unconscious Processes.* 301:3–19.

Bachmann, Talis. 2018. "On a Strategy of Advancement of TMS Based Methods for Studying NCC." *Frontiers in Psychology* 9 (October):2026.

Bachu, Vismaya S., Jayanidhi Kedda, Ian Suk, Jordan J. Green, and Betty Tyler. 2021. "High-Intensity Focused Ultrasound: A Review of Mechanisms and Clinical Applications." *Annals of Biomedical Engineering* 49 (9): 1975–91.

Badran, Bashar W., Kevin A. Caulfield, Sasha Stomberg-Firestein, Philipp M. Summers, Logan T. Dowdle, Matt Savoca, Xingbao Li, et al. 2022. "Sonication of the Anterior Thalamus with MRI-Guided Transcranial Focused Ultrasound (tFUS) Alters Pain Thresholds in Healthy Adults: A Double-Blind, Sham-Controlled Study." *Focus (American Psychiatric Publishing)* 20 (1): 90–99.

Balsdon, Tarryn, and Colin W. G. Clifford. 2018. "Visual Processing: Conscious until Proven Otherwise." *Royal Society Open Science* 5 (1): 171783.

Bareham, Corinne A., Matt Oxner, Tim Gastrell, and David Carmel. 2021. "Beyond the Neural Correlates of Consciousness: Using Brain Stimulation to Elucidate Causal Mechanisms Underlying Conscious States and Contents." *Journal of the Royal Society of New Zealand* 51 (1): 143–70.

Baron, Mark, and Marshall Devor. 2022. "Might Pain Be Experienced in the Brainstem rather than in the Cerebral Cortex?" *Behavioural Brain Research* 427 (113861): 113861.

Barra, Alice, Martin Monti, and Aurore Thibaut. 2022. "Noninvasive Brain Stimulation Therapies to Promote Recovery of Consciousness: Where We Are and Where We Should Go." *Seminars in Neurology* 42 (3): 348–62.

Bault, Nadège, Siti N. Yaakub, and Elsa Fouragnan. 2024. "Early-Phase Neuroplasticity Induced by Offline Transcranial Ultrasound Stimulation in Primates." *Current Opinion in Behavioral Sciences* 56 (101370): 101370.

Bawiec, Christopher R., Peter J. Hollender, Sarah B. Ornellas, Jessica N. Schachtner, Jacob F. Dahill-Fuchel, Soren D. Konecky, and John J. B. Allen. 2025. "A Wearable, Steerable, Transcranial Low-Intensity Focused Ultrasound System." *Journal of Ultrasound in Medicine: Official Journal of the American Institute of Ultrasound in Medicine* 44 (2): 239–61.

Blackmore, Joseph, Shamit Shrivastava, Jerome Sallet, Chris R. Butler, and Robin O. Cleveland. 2019. "Ultrasound Neuromodulation: A Review of Results, Mechanisms and Safety." *Ultrasound in Medicine & Biology* 45 (7): 1509–36.

Block, Ned. 2019. "What Is Wrong with the No-Report Paradigm and How to Fix It." *Trends in Cognitive Sciences*, October. https://doi.org/10.1016/j.tics.2019.10.001.

Block, Ned. 2020. "Finessing the Bored Monkey Problem." *Trends in Cognitive Sciences* 24 (3): 167–68.

Boly, Melanie, Marcello Massimini, Naotsugu Tsuchiya, Bradley R. Postle, Christof Koch, and Giulio Tononi. 2017. "Are the Neural Correlates of Consciousness in the Front or in the Back of the Cerebral Cortex? Clinical and Neuroimaging Evidence." *The Journal of Neuroscience: The Official Journal of the Society for Neuroscience* 37 (40): 9603–13.

Botvinik-Nezer, Rotem, Bogdan Petre, Marta Ceko, Martin A. Lindquist, Naomi P. Friedman, and Tor D. Wager. 2024. "Placebo Treatment Affects Brain Systems Related to Affective and

Cognitive Processes, but Not Nociceptive Pain." *Nature Communications* 15 (1): 6017.
Breitmeyer, Bruno G. 2015. "Psychophysical 'Blinding' Methods Reveal a Functional Hierarchy of Unconscious Visual Processing." *Consciousness and Cognition* 35 (September):234–50.
Cabral-Calderin, Yuranny, and Melanie Wilke. 2020. "Probing the Link between Perception and Oscillations: Lessons from Transcranial Alternating Current Stimulation." *The Neuroscientist: A Review Journal Bringing Neurobiology, Neurology and Psychiatry* 26 (1): 57–73.
Cain, Joshua A., Norman M. Spivak, John P. Coetzee, Julia S. Crone, Micah A. Johnson, Evan S. Lutkenhoff, Courtney Real, et al. 2021. "Ultrasonic Thalamic Stimulation in Chronic Disorders of Consciousness." *Brain Stimulation* 14 (2): 301–3.
Cammalleri, Amanda, Phillip Croce, Wonhye Lee, Kyungho Yoon, and Seung-Schik Yoo. 2020. "Therapeutic Potentials of Localized Blood-Brain Barrier Disruption by Noninvasive Transcranial Focused Ultrasound: A Technical Review: A Technical Review." *Journal of Clinical Neurophysiology: Official Publication of the American Electroencephalographic Society* 37 (2): 104–17.
Chakravarthi, Ramakrishna, and Patrick Cavanagh. 2009. "Recovery of a Crowded Object by Masking the Flankers: Determining the Locus of Feature Integration." *Journal of Vision* 9 (10): 4.1–9.
Chaplin, Vandiver, Marshal A. Phipps, and Charles F. Caskey. 2018. "A Random Phased-Array for MR-Guided Transcranial Ultrasound Neuromodulation in Non-Human Primates." *Physics in Medicine and Biology* 63 (10): 105016.
Choi, Ilsong, Ilayda Demir, Seungmi Oh, and Seung-Hee Lee. 2023. "Multisensory Integration in the Mammalian Brain: Diversity and Flexibility in Health and Disease." *Philosophical Transactions of the Royal Society of London. Series B, Biological Sciences* 378 (1886): 20220338.
Cohen, M. A., K. Ortego, and A. Kyroudis. 2020. "Distinguishing the Neural Correlates of Perceptual Awareness and Postperceptual Processing." *Journal of Neuroscience*. https://www.jneurosci.org/content/40/25/4925.abstract.
Cooke, S. F., and T. V. P. Bliss. 2006. "Plasticity in the Human Central Nervous System." *Brain: A Journal of Neurology* 129 (Pt 7): 1659–73.
Crapse, Trinity B., Hakwan Lau, and Michele A. Basso. 2018. "A Role for the Superior Colliculus in Decision Criteria." *Neuron* 97 (1): 181–94.e6.
Darmani, G., T. O. Bergmann, K. Butts Pauly, C. F. Caskey, L. de Lecea, A. Fomenko, E. Fouragnan, et al. 2022. "Non-Invasive Transcranial Ultrasound Stimulation for Neuromodulation." *Clinical Neurophysiology: Official Journal of the International Federation of Clinical Neurophysiology* 135 (March):51–73.
Dehaene, S. 2014. *Consciousness and the Brain: Deciphering How the Brain Codes Our Thoughts*. Penguin Publishing Group.
Dehaene, Stanislas, and Jean-Pierre Changeux. 2011. "Experimental and Theoretical Approaches to Conscious Processing." *Neuron* 70 (2): 200–227.
Dell'Italia, John, Joseph L. Sanguinetti, Martin M. Monti, Alexander Bystritsky, and Nicco Reggente. 2022. "Current State of Potential Mechanisms Supporting Low Intensity Focused Ultrasound for Neuromodulation." *Frontiers in Human Neuroscience* 16 (April):872639.
Dykstra, Andrew R., Peter A. Cariani, and Alexander Gutschalk. 2017. "A Roadmap for the Study of Conscious Audition and Its Neural Basis." *Philosophical Transactions of the Royal Society of London. Series B, Biological Sciences* 372 (1714). https://doi.org/10.1098/rstb.2016.0103.
Elias, Gavin J. B., Alexandre Boutet, Suresh E. Joel, Jürgen Germann, Dave Gwun, Clemens Neudorfer, Robert M. Gramer, et al. 2021. "Probabilistic Mapping of Deep Brain Stimulation: Insights from 15 Years of Therapy." *Annals of Neurology* 89 (3): 426–43.
Elosegi, Patxi, Ning Mei, and David Soto. 2024. "Characterising the Role of Awareness in Ensemble Perception." *Journal of Experimental Psychology. Human Perception and Performance* 50 (7): 706–22.

Fang, Zepeng, Yuanyuan Dang, An 'an Ping, Chenyu Wang, Qianchuan Zhao, Hulin Zhao, Xiaoli Li, and Mingsha Zhang. 2025. "Human High-Order Thalamic Nuclei Gate Conscious Perception through the Thalamofrontal Loop." *Science (New York, N.Y.)* 388 (6742): eadr3675.

Fan, Joline M., Kai Woodworth, Keith R. Murphy, Leighton Hinkley, Joshua L. Cohen, Joanne Yoshimura, Inhauck Choi, et al. 2024. "Thalamic Transcranial Ultrasound Stimulation in Treatment Resistant Depression." *Brain Stimulation* 17 (5): 1001–4.

Fanselow, Michael S., and Zachary T. Pennington. 2018. "A Return to the Psychiatric Dark Ages with a Two-System Framework for Fear." *Behaviour Research and Therapy* 100 (January):24–29.

Felix, Ciara, Davide Folloni, Haoyu Chen, Jerome Sallet, and Antoine Jerusalem. 2022. "White Matter Tract Transcranial Ultrasound Stimulation, a Computational Study." *Computers in Biology and Medicine* 140 (105094): 105094.

Fleming, Stephen M., Rimona S. Weil, Zoltan Nagy, Raymond J. Dolan, and Geraint Rees. 2010. "Relating Introspective Accuracy to Individual Differences in Brain Structure." *Science* 329 (5998): 1541–43.

Fomenko, Anton, Kai-Hsiang Stanley Chen, Jean-François Nankoo, James Saravanamuttu, Yanqiu Wang, Mazen El-Baba, Xue Xia, et al. 2020. "Systematic Examination of Low-Intensity Ultrasound Parameters on Human Motor Cortex Excitability and Behavior." *eLife* 9 (November). https://doi.org/10.7554/eLife.54497.

Freeman, Daniel K., Donald K. Eddington, Joseph F. Rizzo 3rd, and Shelley I. Fried. 2010. "Selective Activation of Neuronal Targets with Sinusoidal Electric Stimulation." *Journal of Neurophysiology* 104 (5): 2778–91.

Freeman, Daniel K., Jed S. Jeng, Shawn K. Kelly, Espen Hartveit, and Shelley I. Fried. 2011. "Calcium Channel Dynamics Limit Synaptic Release in Response to Prosthetic Stimulation with Sinusoidal Waveforms." *Journal of Neural Engineering* 8 (4): 046005.

Freeman, Daniel K., Joseph F. Rizzo 3rd, and Shelley I. Fried. 2011. "Encoding Visual Information in Retinal Ganglion Cells with Prosthetic Stimulation." *Journal of Neural Engineering* 8 (3): 035005.

Frizon, Leonardo A., Erin A. Yamamoto, Sean J. Nagel, Marian T. Simonson, Olivia Hogue, and Andre G. Machado. 2020. "Deep Brain Stimulation for Pain in the Modern Era: A Systematic Review: A Systematic Review." *Neurosurgery* 86 (2): 191–202.

Fry, F. J., H. W. Ades, and W. J. Fry. 1958. "Production of Reversible Changes in the Central Nervous System by Ultrasound." *Science (New York, N.Y.)* 127 (3289): 83–84.

Graziano, Michael S. A., and Taylor W. Webb. 2015. "The Attention Schema Theory: A Mechanistic Account of Subjective Awareness." *Frontiers in Psychology* 6 (April):500.

Greenspon, E. B., P. Q. Pfordresher, and A. R. Halpern. 2017. "Pitch Imitation Ability in Mental Transformations of Melodies." *Music Perception: An Interdisciplinary Journal* 34 (5): 585–604.

Grippe, Talyta, Yazan Shamli-Oghli, Ghazaleh Darmani, Jean-François Nankoo, Nasem Raies, Can Sarica, Tarun Arora, et al. 2024. "Plasticity-Induced Effects of Theta Burst Transcranial Ultrasound Stimulation in Parkinson's Disease." *Movement Disorders: Official Journal of the Movement Disorder Society* 39 (8): 1364–74.

Halnes, Geir, Tuomo Mäki-Marttunen, Daniel Keller, Klas H. Pettersen, Ole A. Andreassen, and Gaute T. Einevoll. 2016. "Effect of Ionic Diffusion on Extracellular Potentials in Neural Tissue." *PLoS Computational Biology* 12 (11): e1005193.

Hameroff, Stuart, Michael Trakas, Chris Duffield, Emil Annabi, M. Bagambhrini Gerace, Patrick Boyle, Anthony Lucas, Quinlan Amos, Annemarie Buadu, and John J. Badal. 2013. "Transcranial Ultrasound (TUS) Effects on Mental States: A Pilot Study." *Brain Stimulation* 6 (3): 409–15.

IIT-Concerned, Michał Klincewicz, Tony Cheng, Michael Schmitz, Miguel Ángel Sebastián, and Joel S. Snyder. 2025. "What Makes a Theory of Consciousness Unscientific?" *Nature Neuroscience*, March, 1–5.

Jang, Hyunwoo, Panagiotis Fotiadis, George A. Mashour, Anthony G. Hudetz, and Zirui Huang. 2024. "Thalamic Roles in Conscious Perception Revealed by Low-Intensity Focused Ultrasound Neuromodulation." *bioRxivorg*. https://doi.org/10.1101/2024.10.07.617034.

Javid, Ardavan, Sheikh Ilham, and Mehdi Kiani. 2023. "A Review of Ultrasound Neuromodulation Technologies." *IEEE Transactions on Biomedical Circuits and Systems* 17 (5): 1084–96.

Jerusalem, Antoine, Zeinab Al-Rekabi, Haoyu Chen, Ari Ercole, Majid Malboubi, Miren Tamayo-Elizalde, Lennart Verhagen, and Sonia Contera. 2019. "Electrophysiological-Mechanical Coupling in the Neuronal Membrane and Its Role in Ultrasound Neuromodulation and General Anaesthesia." *Acta Biomaterialia* 97 (October):116–40.

Jiménez-Gambín, Sergio, Noé Jiménez, José María Benlloch, and Francisco Camarena. 2019. "Holograms to Focus Arbitrary Ultrasonic Fields through the Skull." *Physical Review Applied* 12 (1). https://doi.org/10.1103/physrevapplied.12.014016.

Jin, J., Pei, G., Ji, Z., Liu, X., Yan, T., Li, W., & Suo, D. (2024). Transcranial focused ultrasound precise neuromodulation: A review of focal size regulation, treatment efficiency and mechanisms. *Frontiers in Neuroscience,* 18, 1463038

Jun, Elizabeth J., Alex R. Bautista, Michael D. Nunez, Daicia C. Allen, J. Tak, Eduardo Alvarez, and M. Basso. 2021. "Causal Role for the Primate Superior Colliculus in the Computation of Evidence for Perceptual Decisions." *Nature Neuroscience* 24 (8): 1121–31.

Kaduk, Kristin, Melanie Wilke, and Igor Kagan. 2024. "Dorsal Pulvinar Inactivation Leads to Spatial Selection Bias without Perceptual Deficit." *Scientific Reports* 14 (1): 12852.

Kastner, Sabine, Ian C. Fiebelkorn, and Manoj K. Eradath. 2020. "Dynamic Pulvino-Cortical Interactions in the Primate Attention Network." *Current Opinion in Neurobiology* 65 (December):10–19.

Kastner, S., Y. Saalmann, and K. Schneider. 2012. "Thalamic Control of Visual Attention." *The Neuroscience of Attention: The Neuroscience of Attention: Attentional Control and Selection*, January, 54–80.

Keogh, Rebecca, and Joel Pearson. 2018. "The Blind Mind: No Sensory Visual Imagery in Aphantasia." *Cortex; a Journal Devoted to the Study of the Nervous System and Behavior* 105 (August):53–60.

Kim, Ho-Jeong, Tien Thuy Phan, Keunhyung Lee, Jeong Sook Kim, Sang-Yeong Lee, Jung Moo Lee, Jongrok Do, et al. 2024. "Long-Lasting Forms of Plasticity through Patterned Ultrasound-Induced Brainwave Entrainment." *Science Advances* 10 (8): eadk3198.

Kim, Hyun-Chul, Wonhye Lee, Daniel S. Weisholtz, and Seung-Schik Yoo. 2023. "Transcranial Focused Ultrasound Stimulation of Cortical and Thalamic Somatosensory Areas in Human." *PloS One* 18 (7): e0288654.

Kim, Young Goo, Song E. Kim, Jihye Lee, Sungeun Hwang, Seung-Schik Yoo, and Hyang Woon Lee. 2022. "Neuromodulation Using Transcranial Focused Ultrasound on the Bilateral Medial Prefrontal Cortex." *Journal of Clinical Medicine* 11 (13): 3809.

Knotts, J. D., B. Odegaard, H. Lau, and D. Rosenthal. 2019. "Subjective Inflation: Phenomenology's Get-Rich-Quick Scheme." *Current Opinion in*. https://www.sciencedirect.com/science/article/pii/S2352250X18301702.

Koch, C. 2018. "What Is Consciousness?" *Nature* 557 (May):S8–12.

Komura, Yutaka, Akihiko Nikkuni, Noriko Hirashima, Teppei Uetake, and Aki Miyamoto. 2013. "Responses of Pulvinar Neurons Reflect a Subject's Confidence in Visual Categorization." *Nature Neuroscience* 16 (6): 749–55.

Ko, Yoshiaki, and Hakwan Lau. 2012. "A Detection Theoretic Explanation of Blindsight Suggests a Link between Conscious Perception and Metacognition." *Philosophical Transactions of the Royal Society of London. Series B, Biological Sciences* 367 (1594): 1401–11.

Kozuch, Benjamin. 2024. "An Embarrassment of Richnesses: The PFC Isn't the Content NCC." *Neuroscience of Consciousness* 2024 (1): niae017.
Kronemer, Sharif I., Mark Aksen, Julia Z. Ding, Jun Hwan Ryu, Qilong Xin, Zhaoxiong Ding, Jacob S. Prince, et al. 2022. "Human Visual Consciousness Involves Large Scale Cortical and Subcortical Networks Independent of Task Report and Eye Movement Activity." *Nature Communications* 13 (1): 7342.
Kuner, R., Kuner, T. (2020). Cellular circuits in the brain and their modulation in acute and chronic pain, Physiol Rev 101: 213-258
Lamme, Victor A. F. 2006. "Towards a True Neural Stance on Consciousness." *Trends in Cognitive Sciences* 10 (11): 494–501.
Lau, Hakwan C. 2008. "Are We Studying Consciousness Yet?" In *Frontiers of Consciousness*. Oxford: Oxford University Press.
Lau, Hakwan C., and Richard E. Passingham. 2006. "Relative Blindsight in Normal Observers and the Neural Correlate of Visual Consciousness." *Proceedings of the National Academy of Sciences of the United States of America* 103 (49): 18763–68.
Lau, Hakwan, and David Rosenthal. 2011. "Empirical Support for Higher-Order Theories of Conscious Awareness." *Trends in Cognitive Sciences* 15 (8): 365–73.
LeDoux, Joseph E., and Daniel S. Pine. 2016. "Using Neuroscience to Help Understand Fear and Anxiety: A Two-System Framework." *The American Journal of Psychiatry* 173 (11): 1083–93.
Lee, Cheng-Chia, Chien-Chen Chou, Fu-Jung Hsiao, Yi-Hsiu Chen, Chun-Fu Lin, Ching-Jen Chen, Syu-Jyun Peng, Hao-Li Liu, and Hsiang-Yu Yu. 2022. "Pilot Study of Focused Ultrasound for Drug-Resistant Epilepsy." *Epilepsia* 63 (1): 162–75.
Lee, Wonhye, Hyun-Chul Kim, Yujin Jung, Yong An Chung, In-Uk Song, Jong-Hwan Lee, and Seung-Schik Yoo. 2016. "Transcranial Focused Ultrasound Stimulation of Human Primary Visual Cortex." *Scientific Reports* 6 (1): 34026.
Lee, Wonhye, Hyungmin Kim, Yujin Jung, In-Uk Song, Yong An Chung, and Seung-Schik Yoo. 2015. "Image-Guided Transcranial Focused Ultrasound Stimulates Human Primary Somatosensory Cortex." *Scientific Reports* 5 (1): 8743.
Lee, Wonhye, Daniel S. Weisholtz, Gary E. Strangman, and Seung-Schik Yoo. 2021. "Safety Review and Perspectives of Transcranial Focused Ultrasound Brain Stimulation." *Brain & Neurorehabilitation* 14 (1): e4.
Legon, Wynn, Leo Ai, Priya Bansal, and Jerel K. Mueller. 2018. "Neuromodulation with Single-Element Transcranial Focused Ultrasound in Human Thalamus." *Human Brain Mapping* 39 (5): 1995–2006.
Legon, Wynn, Tomokazu F. Sato, Alexander Opitz, Jerel Mueller, Aaron Barbour, Amanda Williams, and William J. Tyler. 2014. "Transcranial Focused Ultrasound Modulates the Activity of Primary Somatosensory Cortex in Humans." *Nature Neuroscience* 17 (2): 322–29.
Legon, Wynn, and Andrew Strohman. 2024. "Low-Intensity Focused Ultrasound for Human Neuromodulation" 4 (1): 91.
Macmillan, Neil A., and C. Douglas Creelman. 2005. *Detection Theory: A User's Guide, 2nd Ed*. Lawrence Erlbaum Associates Publishers.
Mahoney, James J., Marc W. Haut, Jeffrey Carpenter, Manish Ranjan, Daisy G. Y. Thompson-Lake, Jennifer L. Marton, Wanhong Zheng, et al. 2023. "Low-Intensity Focused Ultrasound Targeting the Nucleus Accumbens as a Potential Treatment for Substance Use Disorder: Safety and Feasibility Clinical Trial." *Frontiers in Psychiatry* 14 (September):1211566.
Maimbourg, Guillaume, Alexandre Houdouin, Thomas Deffieux, Mickael Tanter, and Jean-Francois Aubry. 2020. "Steering Capabilities of an Acoustic Lens for Transcranial Therapy: Numerical and Experimental Studies." *IEEE Transactions on Bio-Medical Engineering* 67 (1): 27–37.
Malach, R. 2021. "Local Neuronal Relational Structures Underlying the Contents of Human

Conscious Experience." *Neuroscience of Consciousness* 2021 (January). https://doi.org/10.1093/nc/niab028.
Mashour, George A., Pieter Roelfsema, Jean-Pierre Changeux, and Stanislas Dehaene. 2020. "Conscious Processing and the Global Neuronal Workspace Hypothesis." *Neuron* 105 (5): 776–98.
Matt, Eva, Sonja Radjenovic, Michael Mitterwallner, and Roland Beisteiner. 2024. "Current State of Clinical Ultrasound Neuromodulation." *Frontiers in Neuroscience* 18 (June):1420255.
McIntyre, Cameron C., Warren M. Grill, David L. Sherman, and Nitish V. Thakor. 2004. "Cellular Effects of Deep Brain Stimulation: Model-Based Analysis of Activation and Inhibition." *Journal of Neurophysiology* 91 (4): 1457–69.
McNorgan, Chris. 2012. "A Meta-Analytic Review of Multisensory Imagery Identifies the Neural Correlates of Modality-Specific and Modality-General Imagery." *Frontiers in Human Neuroscience* 6 (October):285.
Mehta, Neil, and George A. Mashour. 2013. "General and Specific Consciousness: A First-Order Representationalist Approach." *Frontiers in Psychology* 4 (July):407.
Merker, Bjorn. 2007. "Consciousness without a Cerebral Cortex: A Challenge for Neuroscience and Medicine." *The Behavioral and Brain Sciences* 30 (1): 63–81; discussion 81–134.
Merker, B. 2013. "The Efference Cascade, Consciousness, and Its Self: Naturalizing the First Person Pivot of Action Control." *Frontiers in Psychology* 4 (August):501.
Michel, M. 2022. "Conscious Perception and the Prefrontal Cortex A Review." *Journal of Consciousness Studies: Controversies in Science & the Humanities*, July. https://doi.org/10.53765/20512201.29.7.115.
Michel, Matthias, Diane Beck, Ned Block, Hal Blumenfeld, Richard Brown, David Carmel, Marisa Carrasco, et al. 2019. "Opportunities and Challenges for a Maturing Science of Consciousness." *Nature Human Behaviour* 3 (2): 104–7.
Monti, Martin. 2021. "Ultrasonic Stimulation in Disorders of Consciousness." *Brain Stimulation* 14 (6): 1746–47.
Morales, Jorge, and Hakwan Lau. n.d. "Confidence Tracks Consciousness." In *Qualitative Consciousness: Themes from the Philosophy of David Rosenthal*, edited by Joshua Weisberg.
Morales, Jorge, Brian Odegaard, and Brian Maniscalco. 2022. "10 The Neural Substrates of Conscious Perception without Performance Confounds." In *Neuroscience and Philosophy*, edited by Felipe De Brigard and Walter Sinnott-Armstrong. Cambridge (MA): MIT Press.
Mudrik, Liad, Melanie Boly, Stanislas Dehaene, Stephen M. Fleming, Victor Lamme, Anil Seth, and Lucia Melloni. 2025. "Unpacking the Complexities of Consciousness: Theories and Reflections." *Neuroscience and Biobehavioral Reviews* 170 (106053): 106053.
Nakajima, Koji, Takahiro Osada, Akitoshi Ogawa, Masaki Tanaka, Satoshi Oka, Koji Kamagata, Shigeki Aoki, Yasushi Oshima, Sakae Tanaka, and Seiki Konishi. 2022. "A Causal Role of Anterior Prefrontal-Putamen Circuit for Response Inhibition Revealed by Transcranial Ultrasound Stimulation in Humans." *Cell Reports* 40 (7): 111197.
Nandi, Tulika, Ainslie Johnstone, Eleanor Martin, Catharina Zich, Robert Cooper, Sven Bestmann, Til Ole Bergmann, Bradley Treeby, and Charlotte J. Stagg. 2023. "Ramped V1 Transcranial Ultrasonic Stimulation Modulates but Does Not Evoke Visual Evoked Potentials." *Brain Stimulation* 16 (2): 553–55.
Nandi, Tulika, Benjamin R. Kop, Kim Butts Pauly, Charlotte J. Stagg, and Lennart Verhagen. 2024. "The Relationship between Parameters and Effects in Transcranial Ultrasonic Stimulation." *Brain Stimulation* 17 (6): 1216–28.
Nieder, Andreas. 2021. "Consciousness without Cortex." *Current Opinion in Neurobiology* 71 (December):69–76.
Odegaard, Brian, Min Yu Chang, Hakwan Lau, and Sing-Hang Cheung. 2018. "Inflation versus Filling-in: Why We Feel We See More than We Actually Do in Peripheral Vision."

    *Philosophical Transactions of the Royal Society of London. Series B, Biological Sciences* 373 (1755): 20170345.
Odegaard, Brian, Robert T. Knight, and Hakwan Lau. 2017. "Should a Few Null Findings Falsify Prefrontal Theories of Conscious Perception?" *The Journal of Neuroscience: The Official Journal of the Society for Neuroscience* 37 (40): 9593–9602.
Oh, S. J., Lee, J. M., Kim, H. B., Lee, J., Han, S., Bae, J. Y., et al. (2019). Ultrasonic neuromodulation via astrocytic *TRPA1*. *Current Biology*, 29(20), 3386–3401.e8.
Pellow, Carly, Samuel Pichardo, and G. Bruce Pike. 2024. "A Systematic Review of Preclinical and Clinical Transcranial Ultrasound Neuromodulation and Opportunities for Functional Connectomics." *Brain Stimulation* 17 (4): 734–51.
Persaud, Navindra, Matthew Davidson, Brian Maniscalco, Dean Mobbs, Richard E. Passingham, Alan Cowey, and Hakwan Lau. 2011. "Awareness-Related Activity in Prefrontal and Parietal Cortices in Blindsight Reflects More than Superior Visual Performance." *NeuroImage* 58 (2): 605–11.
Persaud, Navindra, Peter McLeod, and Alan Cowey. 2007. "Post-Decision Wagering Objectively Measures Awareness." *Nature Neuroscience* 10 (2): 257–61.
Peters, Megan A. K., and Hakwan Lau. 2015. "Human Observers Have Optimal Introspective Access to Perceptual Processes Even for Visually Masked Stimuli." *eLife* 4 (October):e09651.
Pincham, Hannah L., H. Bowman, and D. Szucs. 2016. "The Experiential Blink: Mapping the Cost of Working Memory Encoding onto Conscious Perception in the Attentional Blink." *Cortex; a Journal Devoted to the Study of the Nervous System and Behavior* 81 (August):35–49.
Plaksin, Michael, Eitan Kimmel, and Shy Shoham. 2016. "Cell-Type-Selective Effects of Intramembrane Cavitation as a Unifying Theoretical Framework for Ultrasonic Neuromodulation." *eNeuro* 3 (3): ENEURO.0136–15.2016.
Pounder, Zoë, Jane Jacob, Samuel Evans, Catherine Loveday, Alison F. Eardley, and Juha Silvanto. 2022. "Only Minimal Differences between Individuals with Congenital Aphantasia and Those with Typical Imagery on Neuropsychological Tasks That Involve Imagery." *Cortex; a Journal Devoted to the Study of the Nervous System and Behavior* 148 (March):180–92.
Raccah, Omri, Ned Block, and Kieran C. R. Fox. 2021. "Does the Prefrontal Cortex Play an Essential Role in Consciousness? Insights from Intracranial Electrical Stimulation of the Human Brain." *The Journal of Neuroscience: The Official Journal of the Society for Neuroscience* 41 (10): 2076–87.
Rahnev, Dobromir, Brian Maniscalco, Tashina Graves, Elliott Huang, Floris P. de Lange, and Hakwan Lau. 2011. "Attention Induces Conservative Subjective Biases in Visual Perception." *Nature Neuroscience* 14 (12): 1513–15.
Recht, Samuel, P. Mamassian, and Vincent de Gardelle. 2019. "Temporal Attention Causes Systematic Biases in Visual Confidence." *Scientific Reports* 9 (August). https://doi.org/10.1038/s41598-019-48063-x.
Redinbaugh, Michelle J., and Yuri B. Saalmann. 2024. "Contributions of Basal Ganglia Circuits to Perception, Attention, and Consciousness." *Journal of Cognitive Neuroscience* 36 (8): 1620–42.
Riis, Thomas S., Daniel A. Feldman, Adam J. Losser, Akiko Okifuji, and Jan Kubanek. 2024. "Noninvasive Targeted Modulation of Pain Circuits with Focused Ultrasonic Waves." *Pain* 165 (12): 2829–39.
Rounis, Elisabeth, Brian Maniscalco, John C. Rothwell, Richard E. Passingham, and Hakwan Lau. 2010. "Theta-Burst Transcranial Magnetic Stimulation to the Prefrontal Cortex Impairs Metacognitive Visual Awareness." *Cognitive Neuroscience* 1 (3): 165–75.

Saalmann, Yuri B., and Sabine Kastner. 2011. "Cognitive and Perceptual Functions of the Visual Thalamus." *Neuron* 71 (2): 209–23.
Sanguinetti, Joseph L., Stuart Hameroff, Ezra E. Smith, Tomokazu Sato, Chris M. W. Daft, William J. Tyler, and John J. B. Allen. 2020. "Transcranial Focused Ultrasound to the Right Prefrontal Cortex Improves Mood and Alters Functional Connectivity in Humans." *Frontiers in Human Neuroscience* 14 (February):52.
Schnakers, Caroline, Alex Korb, Alexander Bystritsky, Paul Vespa, and Martin Monti. 2018. "Low Intensity Focused Ultrasound as a Treatment for Disorders of Consciousness: Preliminary Data." *Archives of Physical Medicine and Rehabilitation* 99 (10): e41–42.
Sekimoto, Taisei, and Isamu Motoyoshi. 2022. "Ensemble Perception without Phenomenal Awareness of Elements." *Scientific Reports* 12 (July). https://doi.org/10.1038/s41598-022-15850-y.
Seth, Anil K., and Tim Bayne. 2022. "Theories of Consciousness." *Nature Reviews. Neuroscience* 23 (7): 439–52.
Shin, Minwoo, Minjee Seo, Seung-Schik Yoo, and Kyungho Yoon. 2024. "TFUSFormer: Physics-Guided Super-Resolution Transformer for Simulation of Transcranial Focused Ultrasound Propagation in Brain Stimulation." *IEEE Journal of Biomedical and Health Informatics* 28 (7): 4024–35.
Sorum, B., Rietmeijer, R.A., Gopakumar, K., Brohawn, S.G. (2021). Ultrasound activates mechanosensitive TRAAK K+ channels through the lipid membrane, *Proceedings of the National Academy of Sciences*, 118(6)
Spagna, Alfredo, Zoe Heidenry, Michelle Miselevich, Chloe Lambert, Benjamin E. Eisenstadt, Laura Tremblay, Zixin Liu, Jianghao Liu, and Paolo Bartolomeo. 2024. "Visual Mental Imagery: Evidence for a Heterarchical Neural Architecture." *Physics of Life Reviews* 48 (March):113–31.
Storm, Johan F., P. Christiaan Klink, Jaan Aru, Walter Senn, Rainer Goebel, Andrea Pigorini, Pietro Avanzini, et al. 2024. "An Integrative, Multiscale View on Neural Theories of Consciousness." *Neuron* 112 (10): 1531–52.
Striedter, G. F. 2005. "Principles of Brain Evolution." https://psycnet.apa.org/record/2004-21314-000.
Tian, Xing, and David Poeppel. 2010. "Mental Imagery of Speech and Movement Implicates the Dynamics of Internal Forward Models." *Frontiers in Psychology* 1 (October):166.
Tononi, Giulio. 2012. *Phi: A Voyage from the Brain to the Soul*. New York, NY: Pantheon.
Tufail, Yusuf, Alexei Matyushov, Nathan Baldwin, Monica L. Tauchmann, Joseph Georges, Anna Yoshihiro, Stephen I. Helms Tillery, and William J. Tyler. 2010. "Transcranial Pulsed Ultrasound Stimulates Intact Brain Circuits." *Neuron* 66 (5): 681–94.
Tyler, William J., Yusuf Tufail, Michael Finsterwald, Monica L. Tauchmann, Emily J. Olson, and Cassondra Majestic. 2008. "Remote Excitation of Neuronal Circuits Using Low-Intensity, Low-Frequency Ultrasound." *PloS One* 3 (10): e3511.
Usrey, W. M., and S. Kastner. 2020. "Functions of the Visual Thalamus in Selective Attention." *The Cognitive Neurosciences*. https://books.google.com/books?hl=en&lr=&id=h63jDwAAQBAJ&oi=fnd&pg=PA367&dq=kastner+thalamus+attention&ots=thXoeIqGvN&sig=PtyFBL99rI1U7apSXDxSCNLU0FM.
Verhagen, Lennart, Cécile Gallea, Davide Folloni, Charlotte Constans, Daria Ea Jensen, Harry Ahnine, Léa Roumazeilles, et al. 2019. "Offline Impact of Transcranial Focused Ultrasound on Cortical Activation in Primates." *eLife* 8 (February):e40541.
Voytek, Bradley, Matar Davis, Elena Yago, Francisco Barceló, Edward K. Vogel, and Robert T. Knight. 2010. "Dynamic Neuroplasticity after Human Prefrontal Cortex Damage." *Neuron* 68 (3): 401–8.
Weber, Simon, Thomas Christophel, Kai Görgen, Joram Soch, and John-Dylan Haynes. 2024.

"Working Memory Signals in Early Visual Cortex Are Present in Weak and Strong Imagers." *Human Brain Mapping* 45 (3): e26590.

Weiskrantz, Lawrence. 2009. "Blindsight: A Case Study Spanning 35 Years and New Developments." Oxford University PressOxford. https://doi.org/10.1093/oso/9780199567218.001.0001.

Yaakub, Siti N., Tristan A. White, Jamie Roberts, Eleanor Martin, Lennart Verhagen, Charlotte J. Stagg, Stephen Hall, and Elsa F. Fouragnan. 2023. "Transcranial Focused Ultrasound-Mediated Neurochemical and Functional Connectivity Changes in Deep Cortical Regions in Humans." *Nature Communications* 14 (1): 5318.

Yoon, Kyungho, Wonhye Lee, Phillip Croce, Amanda Cammalleri, and Seung-Schik Yoo. 2018. "Multi-Resolution Simulation of Focused Ultrasound Propagation through Ovine Skull from a Single-Element Transducer." *Physics in Medicine and Biology* 63 (10): 105001.

Yoo, S.S., Bystritsky, A., Lee, J.H., Zhang, Y., Fischer, K., Min, B.K., McDonnaold, N., Pascual-Leone, A., Jolesz, F.A. (2011), Neuroimage, Jun 1; 56(3): 1267-75

Yoo, S.H., Croce, P., Margolin, R.W., Lee, S.D., Lee, W. (2017). Pulsed focused ultrasound chnages in nerve conduction of earthworm giant axonal fibers, *Neuroreport*, Mar 1; 28(4): 229-233

Yoo, S.-S., Yoon, K., Croce, P., Cammalleri, A., Margolin, R. W., & Lee, W. (2018). Focused ultrasound brain stimulation to anesthetized rats induces long-term changes in somatosensory evoked potentials. *International Journal of Imaging Systems and Technology*, **28**(2), 106–112

Yoo, Sangjin, David R. Mittelstein, Robert C. Hurt, Jerome Lacroix, and Mikhail G. Shapiro. 2022. "Focused Ultrasound Excites Cortical Neurons via Mechanosensitive Calcium Accumulation and Ion Channel Amplification." *Nature Communications* 13 (1): 493.

Zeki, Semir. 2008. "The Disunity of Consciousness." *Progress in Brain Research* 168:11–18.

Zhang, Tingting, Na Pan, Yuping Wang, Chunyan Liu, and Shimin Hu. 2021. "Transcranial Focused Ultrasound Neuromodulation: A Review of the Excitatory and Inhibitory Effects on Brain Activity in Human and Animals." *Frontiers in Human Neuroscience* 15 (September):749162.

Zhang, H., Wang, D. Wei, P., Fan, X., Yang, Y., An, Y., Dai, Y., Feng, T., Shan, Y., Ren, L., Zzhao, G. (2023). Integrative roles of human amygdala subdivisions: insight from direct intracerebral stimulations via stereotactic EEG, *Human Brain Mapping*, 44:3610-3623

Zheng, Jieyu. 2024. "Mice in the Manhattan Maze: Rapid Learning, Flexible Routing and Generalization, with and without Cortex." 2024. https://2024.ccneuro.org/pdf/320_Paper_authored_CCN_2024_wtitle.pdf.